\newif\ifarxiv
\arxivtrue

\newif\ifreview
\reviewfalse

\ifreview
\documentclass[sigconf,nonacm,screen,natbib=false,review,anonymous,balance=false]{acmart}
\else
    \ifarxiv
    \documentclass[sigconf,nonacm,screen,natbib=false]{acmart}
    \else
    \documentclass[sigconf,screen,natbib=false]{acmart}
    \fi
\fi

\usepackage{subcaption}
\usepackage{placeins}

\usepackage{microtype}
\usepackage{enumitem}

\usepackage{xspace}

\usepackage{siunitx}
\DeclareSIUnit\syear{yr}

\usepackage{algorithm}
\usepackage{algpseudocodex}
\usepackage{amsmath}
\usepackage{makecell}
\usepackage{nicefrac}

\ifreview
\usepackage[font=small,skip=0pt]{caption}
\fi

\usepackage{enumitem}
\setlist{noitemsep,topsep=0pt,parsep=0pt,partopsep=0pt,leftmargin=*}

\usepackage[table,dvipsnames]{xcolor}

\definecolor{babyblueeyes}{rgb}{0.63, 0.79, 0.95}

\definecolor{boldblue}{HTML}{3A5DAE} 
\definecolor{newhorizon}{HTML}{E04F39} 

\usepackage{pifont}

\AtBeginDocument{%
  }

\setcopyright{acmlicensed}
\copyrightyear{2026}
\acmYear{2026}
\acmDOI{XXXXXXX.XXXXXXX}
\acmConference[Conference acronym 'XX]{Make sure to enter the correct
  conference title from your rights confirmation email}{June 03--05,
  2018}{Woodstock, NY}
\acmISBN{978-1-4503-XXXX-X/2018/06}

\acmSubmissionID{\#133}

\RequirePackage[
  datamodel=acmdatamodel,
  style=acmnumeric,
  sortcites=true,
]{biblatex}

\newcommand{\noind}[0]{\par \noindent}
\newcommand{\noindpar}[1]{\noind {\bf #1}}

\definecolor{blurple}{HTML}{5865F2}

\begin{document}

\ifarxiv
\title{Magnetic Tunnel Junctions for Timekeeping in Intermittent Computing Systems}
\else
\title{``Failure Is an Option'': Broken Memory as a Timekeeper for Intermittent Computing}
\fi

\newcommand{\sysname}{FLINT\xspace}

\author{Nikola Vuk Maruszewski}
\email{nikola@gatech.edu}
\orcid{0009-0009-5468-4085}
\affiliation{%
  \institution{Georgia Institute of Technology}
  \city{Atlanta}
  \state{GA}
  \country{USA}
}

\author{Jordan Athas}
\email{jordan.athas@northwestern.edu}
\orcid{0009-0005-9209-2786}
\affiliation{%
  \institution{Northwestern University}
  \city{Evanston}
  \state{IL}
  \country{USA}
}

\author{Allison Fleming}
\email{allisonfleming2030@u.northwestern.edu}
\orcid{0009-0001-7114-3311}
\affiliation{%
  \institution{Northwestern University}
  \city{Evanston}
  \state{IL}
  \country{USA}
}

\author{Christian Duffee}
\email{christian.duffee@northwestern.edu}
\orcid{0000-0001-7249-7122}
\affiliation{%
  \institution{Northwestern University}
  \city{Evanston}
  \state{IL}
  \country{USA}
}

\author{Eren Yildiz}
\email{eyildiz8@gatech.edu}
\orcid{0000-0002-4631-7834}
\affiliation{%
  \institution{Georgia Institute of Technology}
  \city{Atlanta}
  \state{GA}
  \country{USA}
}

\author{Saad Ahmed}
\email{sahmed368@gatech.edu}
\orcid{0000-0002-0341-2997}
\affiliation{%
  \institution{Georgia Institute of Technology}
  \city{Atlanta}
  \state{GA}
  \country{USA}
}

\author{Yaman Sangar}
\email{ysangar3@gatech.edu}
\orcid{0000-0002-7056-961X}
\affiliation{%
  \institution{Georgia Institute of Technology}
  \city{Atlanta}
  \state{GA}
  \country{USA}
}

\author{Pedram Khalili}
\email{pedram@northwestern.edu}
\orcid{0000-0002-1539-1521}
\affiliation{%
  \institution{Northwestern University}
  \city{Evanston}
  \state{IL}
  \country{USA}
}

\author{Josiah Hester}
\email{josiah@gatech.edu}
\orcid{0000-0002-1680-085X}
\affiliation{%
  \institution{Georgia Institute of Technology}
  \city{Atlanta}
  \state{GA}
  \country{USA}
}

\renewcommand{\shortauthors}{Maruszewski et al.}

\begin{abstract}
Batteryless intermittent systems run unattended for years, but power failures erase timekeeping state, corrupting sensing, scheduling, and coordination. State-of-the-art timekeepers infer elapsed time from capacitor discharge; however, the capacitor must be sized for the longest interval measured (so range, energy, and area grow together), and repeated charge--discharge cycling lowers capacitance over time, biasing every estimate further as the deployment ages. We present \sysname, a timekeeper that reads elapsed time from the stochastic retention loss of an array of ``broken'' Magnetic Tunnel Junctions (MTJs)---spintronic memory cells engineered to lose state predictably. Because the decay timescale is fixed by device geometry, the energy to read it is independent of the interval measured and does not drift with device age. We validate \sysname's array model against 21 fabricated MTJs, then evaluate the full timekeeper in real-device-trace-driven simulation, showing that it tracks \textbf{over 15 minutes} of off-time within 10\% error while consuming \textbf{only \qty[detect-weight]{1.03}{\micro\joule}} and occupying \textbf{under \qty[detect-weight]{0.1}{\mm^2}}---\textbf{9.2\texttimes} the range at \textbf{11\texttimes} lower energy than prior work. It extends to longer intervals at no added cost, and makes 16--52\texttimes{} fewer scheduling errors than an aging capacitor clock over a one-year deployment.
\end{abstract}

\ifarxiv
\received{\today}
\else
\received{05 June 2026}
\fi

\maketitle

\section{Introduction}
\begin{figure}
    \centering \includegraphics[width=0.9\linewidth]{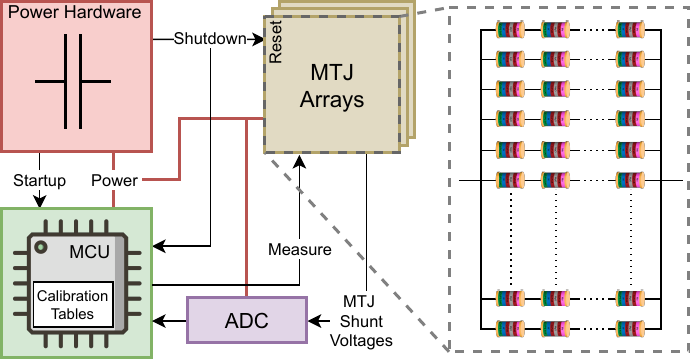}
    \caption{\sysname timekeeper system overview and dataflow. Instead of capacitive decay, \sysname relies on broken memory (MTJ arrays) tuned to lose data predictably during power failures. On power return, a simple resistance-based array readout measures elapsed time.}
    \Description{Block diagram showing FLINT system architecture with power management, MTJ arrays, and readout circuitry for time measurement.}
    \label{fig:sys_overview}
\end{figure}



Liberating embedded sensing devices from batteries and powering them entirely with energy harvested from ambient sources---solar~\cite{geissdoerfer2021bootstrapping}, RF~\cite{menon2022wireless}, and thermal~\cite{bakar2022protean}---has unlocked applications previously out of reach~\cite{curtiss2021facebit,cheng2024eagleeye,de2020battery,desai2022camaroptera}. 
Harvested energy accumulates in a small capacitor; the device runs only while this energy is available and powers down once it is depleted~\cite{doglioni2025capdyn,williams2024energy}, repeatedly cycling between execution and complete power loss. 
This makes frequent power failures the norm for these devices, and so computing becomes \textit{intermittent}. 
To ensure forward progress, intermittent computing runtimes preserve computation state~\cite{ransford2012mementos,balsamo2016hibernus++,ahmed2019efficient}, memory consistency~\cite{maeng2017alpaca,yildirim2018ink}, and peripheral context~\cite{yildiz2023efficient,branco2019intermittent} across power failures, allowing execution to resume once energy returns.



With no batteries to service, these systems can be used in long-term \textit{unattended} deployments.
However, while existing runtimes preserve system state, they cannot preserve temporal continuity, yet many intermittent applications require time-sensitive operations such as periodic sensing, timeout management, and event scheduling~\cite{hester2017timely,islam2020scheduling,kortbeek2020time,maeng2020adaptive,erata2023etap,yildiz2024adaptable,yildiz2023efficient, hester_TimelyExecutionIntermittently_2017}. For example, wildlife tracking collars periodically record GPS locations every few seconds~\cite{juang2002zebranet}, cold-chain monitoring systems sample temperature every few minutes to detect spoilage~\cite{mercier2017coldchain}, and infrastructure monitoring platforms schedule strain measurements days apart~\cite{afanasov2020battery}. Consequently, the required timekeeping interval varies by orders of magnitude across deployments, ranging from seconds to minutes to days. In all cases, losing track of elapsed time can result in missed sensing events, premature actions, stale data, or incorrect timeout decisions that can compromise system correctness for the lifetime of an unattended deployment.
Unpredictable ambient energy availability further exacerbates this problem, as it causes power interruptions to vary significantly in timing and duration.

Together, these factors make reliable timekeeping a fundamental requirement for batteryless systems.
We therefore identify three requirements for a practical batteryless timekeeper:
\begin{itemize}[leftmargin=0.22in]

\item[\textbf{R1}] \textbf{Application-agnostic operation.}
\textit{A timekeeping solution must be generalizable, not bespoke}---independent of application-specific assumptions, workloads, energy sources, or deployment environments. Its energy and latency overhead must remain flat across all operating ranges---short intervals to multi-hour gaps alike---without hardware modifications or re-sizing.

\item[\textbf{R2}] \textbf{Temporal stability.}
\textit{A timekeeper must stay accurate across the full operational lifetime} of the device, not just at deployment. Component aging causes slow, cumulative drift---despite rated accuracy at deployment, the inaccuracies will only increase over time---and correcting it requires recalibration, which is impractical in most remote deployments.

\item[\textbf{R3}] \textbf{Environmental robustness.} \textit{A timekeeper must remain accurate despite fast-changing conditions} (e.g., supply voltage variation, temperature swings) that shift its readings within a single power cycle. Unlike aging, these perturbations are immediate and reversible, but no less damaging if uncompensated.

\end{itemize}



\begin{figure}
    \centering
    \includegraphics[width=0.9\linewidth]{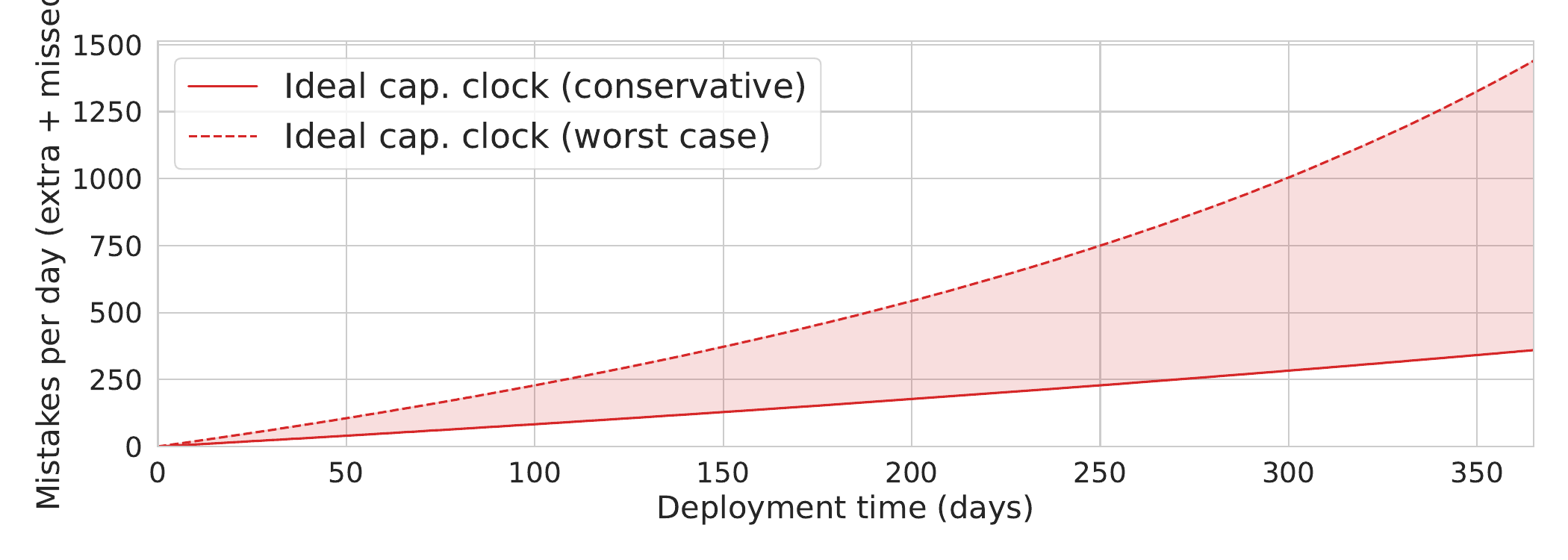}
    \caption{Daily capacitor clock mistakes over 1 yr. Bands: conservative (\qty{20}{\percent}/yr) to worst-case (\qty{50}{\percent}/yr) degradation. See \autoref{sec:eval-aging}, \autoref{sec:results-aging}.}
    \Description{Graph showing accumulated sampling errors from capacitor aging over one year, with two bands representing conservative and worst-case degradation rates.}
    \label{fig:aging_expirations_cap_only}
\end{figure}

Existing capacitor-based timekeeping approaches~\cite{deep_HARCHeterogeneousArray_2020,dewinkel_ReliableTimekeepingIntermittent_2020,hester2016persistent, arreola2022federated} fail to meet all three requirements. First, they assume capacitor discharge is a stable process. Charge--discharge cycling can degrade capacitance by up to half within a year~\cite{choi2022capos}, which shifts the discharge curve and consistently increases the error in every future estimate (\textbf{R2}). \autoref{fig:aging_expirations_cap_only} shows missed and redundant samples accumulating as the capacitor ages. 
This drift is directional and cannot be corrected without recalibration, making capacitor clocks a futile attempt for unattended multi-year deployments. Furthermore, range and cost are correlated: a capacitor must be sized for the longest measurement interval and recharged each cycle; range, energy, area, and startup latency grow in lockstep~\cite{dewinkel_ReliableTimekeepingIntermittent_2020,deep_HARCHeterogeneousArray_2020,geissdoerfer2024riotee}. Reaching a range of 15 minutes requires $\approx\!\qty{101}{\uJ}$ of energy (\autoref{fig:energy_vs_range})---on the same order of magnitude as some main system capacitors~\cite{geissdoerfer2022learning}. 


Second, capacitor-based timekeepers must be calibrated to specific workloads, harvesting conditions, and discharge profiles, limiting generality (\textbf{R1}). Larger off-time ranges demand larger capacitors, which charge more slowly, thereby increasing startup latency and reducing the fraction of events the device can capture. Lastly, capacitor discharge shifts with supply voltage and temperature, requiring recalibration after any environmental change (\textbf{R3}); correcting aging-induced drift demands recalibration too (\textbf{R2}). Both are infeasible for devices deployed for years in hard-to-reach areas.

\noindent{\textbf{Contributions:}} We take a fundamentally different approach: measuring time from device physics rather than stored energy. We leverage MTJs, spintronic memory cells engineered to ``fail'' at retaining data in a controlled, stochastic manner. Unlike conventional non-volatile memory, where retention loss is a defect, these predictable failures produce a stable statistical decay across an MTJ array. Reading this decay reveals elapsed time directly from device behavior, decoupling timekeeping cost from the interval measured.

We introduce Faulty Logic for Intermittent Nanojoule Timekeeping (\sysname), comprising MTJ devices, supporting circuitry, and a lightweight runtime kernel. FLINT (\autoref{fig:sys_overview}) organizes MTJs into arrays that generate a stable decay signal (\textbf{R2}) mapped to elapsed time. Time estimation uses simple ADC measurements with lightweight calibration and temperature correction (\textbf{R3}). \sysname tracks \qty{938}{\s} (over 15 minutes) of off-time within 10\% geomean error while consuming \qty{1.03}{\micro\joule}. It occupies less than \qty{0.1}{\mm^2} and adds \qty{50}{\us} of latency to reset arrays at power-off and \qty{70.83}{\us} to estimate time at power-on. This range (\qty{938}{\s}) is not a hard limit: an array's range is set by its device geometry, not stored energy, so \sysname extends to longer intervals without spending more energy or area---something a capacitor cannot do. Arrays draw no power when idle and occupy little area, so a single chip can hold many and select the appropriate range at runtime (\textbf{R1}). \sysname makes the following contributions:

\begin{itemize}
    \item \textbf{Aging as an overlooked failure mode.} We identify capacitor aging as a critical but largely ignored source of timekeeping error in long-lived batteryless deployments, and quantify its effect: an aging capacitor clock accumulates $16$--$52\times$ more scheduling errors than \sysname over a one-year deployment.

    \item \textbf{A physics-based timekeeping primitive.} We show that broken MTJs reveal elapsed time through stochastic retention loss, giving the first timekeeper whose energy cost is independent of the measured duration and whose accuracy is immune to charge-cycle aging by construction.

    \item \textbf{The \sysname system.} We realize this primitive with multiple MTJ arrays, simple control hardware, lightweight calibration, temperature correction, and estimate fusion. Idle arrays consume no energy and occupy minimal area, so one chip can host more arrays than needed and let applications select their range at runtime---a ``one-size-fits-all,'' near-instant startup timekeeper.

    \item \textbf{Hardware-grounded evaluation.} We validate our array model against 21 fabricated MTJs and, in simulation (built and validated using fabricated MTJs and modeled circuits), show \sysname tracks over 15 minutes within 10\% error at \qty{1.03}{\micro\joule}, holds its range as it ages, and remains accurate across temperatures. Five application case studies span two families: three accuracy studies (GPS tracking, structural health monitoring, cold-chain logistics) that quantify scheduling mistakes under capacitor aging, and two long-duration scenarios that capacitor clocks cannot cover at all.
\end{itemize}

\section{Background and Motivation}
\begin{figure}[!t]
    \centering
    \includegraphics[width=\linewidth]{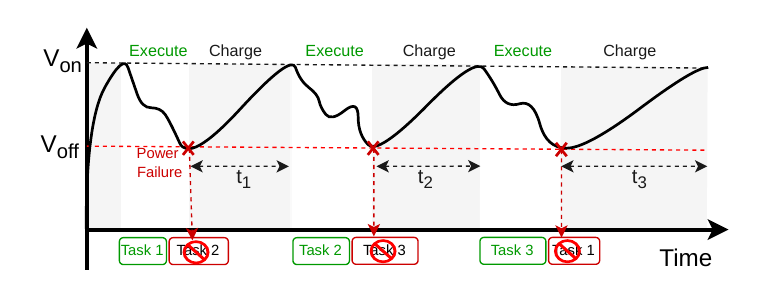}
    \caption{Intermittent system power curve. Note off time variability.
    }
    \Description{Voltage waveform showing intermittent power cycling with varying on and off periods, highlighting time loss during power failures.}
    \label{fig:intpowercurve}
\end{figure}

Batteryless energy-harvesting devices operate on intermittent power cycles, utilizing harvested energy to drive computation whenever it becomes available. As shown in \autoref{fig:intpowercurve}, energy accumulates in a storage capacitor. The device turns on once the voltage exceeds a predefined threshold ($V_{\textit{on}}$), executes until the stored energy is depleted, and turns off when the voltage drops below the minimum operating threshold ($V_{\textit{off}}$). Due to variability in environmental energy sources, both charging time and execution duration are highly irregular, ranging from milliseconds to several minutes. As a result, application execution is repeatedly interrupted across multiple power cycles. This intermittent behavior introduces fundamental challenges for time-sensitive operations. Data collected in one execution phase may become outdated by the time it is processed or transmitted in a later cycle, as the elapsed real time between these phases is unknown. Without a notion of time across power failures, the system cannot determine data freshness or enforce temporal constraints. Consequently, persistent timekeeping becomes essential to support correct and energy-efficient execution of time-dependent tasks.
\subsection{Persistent Timekeeping}
Existing methods for persistent timekeeping can be categorized into three main types: RTCs, software-based, and capacitor-based methods. RTCs provide accurate timekeeping but require a continuous power source, limiting their applicability in fully batteryless settings. Software-based approaches infer time from system behavior, such as SRAM data remanence~\cite{hester_PersistentClocksBatteryless_2016}, oscillator stabilization~\cite{alsubhi2020can}, or ambient power levels~\cite{yildiz2021persistent}, but are often sensitive to environmental conditions~\cite{yildiz2021persistent} and offer limited measurement ranges~\cite{hester_PersistentClocksBatteryless_2016,alsubhi2020can}. 

\begin{figure}
    \centering
    \includegraphics[width=\linewidth]{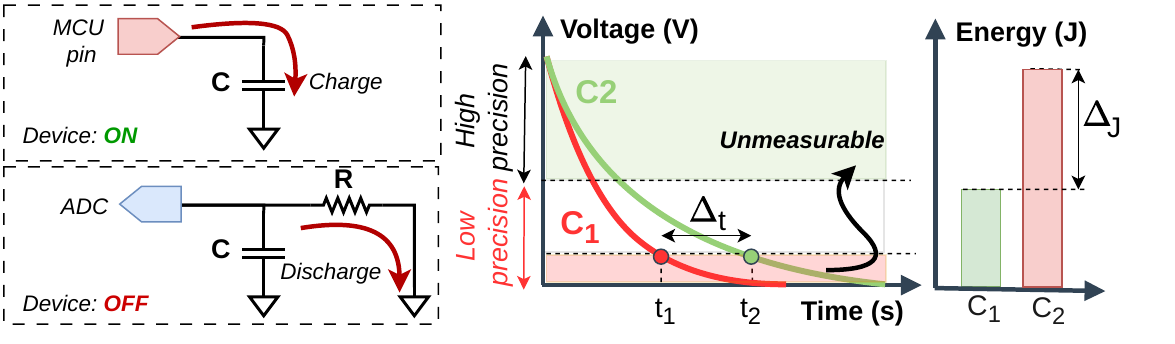}
    \caption{
    Capacitor-based timekeeping charges a capacitor during the on-time and discharges it during the off-time, with measurable duration and energy consumption determined by capacitor size.}
    \Description{Diagram showing capacitor charge and discharge cycles during device on and off periods, with voltage levels used to infer elapsed time.}
    \label{fig:cap_discharge}
\end{figure}

Consequently, many systems utilize capacitor-based timekeeping~\cite{dewinkel_ReliableTimekeepingIntermittent_2020, deep_HARCHeterogeneousArray_2020,arreola2022federated,yildiz_TrackingTimeBetter_2024}, which estimates elapsed time during power failures by monitoring the discharge of dedicated capacitors. In this approach, a capacitor is charged before a power failure and then allowed to discharge while the system is off, as shown in~\autoref{fig:cap_discharge}. When power is restored, the voltage level in the timekeeping capacitor is measured via an ADC on an MCU to calculate the elapsed time using known discharge time characteristics, as shown in \autoref{eq:discharge_time}.
\begin{equation}
    t_{dchrg} = R C \ln\left(\nicefrac{V_{init}}{V_{discharge}}\right)
    \label{eq:discharge_time}
\end{equation}

\( V_{init} \) represents the initial voltage before power failure, and \( V_{discharge} \) is the voltage across the timekeeping capacitor when the system reboots. The measurable time range is highly dependent on the capacitor size; larger capacitors can measure longer durations but require more energy, while smaller capacitors consume less energy but reduce the measurable time range. Additionally, unpredictable power conditions add noise and reduce accuracy.

Furthermore, larger capacitors increase the start-up time following each power failure, as the system must recharge the timekeeping capacitor to prepare for the next outage. Yet larger capacitors are exactly what low-power conditions demand to measure longer durations. Multiple-capacitor schemes~\cite{dewinkel_ReliableTimekeepingIntermittent_2020,deep_HARCHeterogeneousArray_2020,arreola2022federated} add flexibility by combining capacitors of different sizes, but they do not escape the trade-off: every additional second of range is paid for in energy, board area, and startup latency, and costs grow with target range.

A second limitation is that the capacitor degrades with use. Repeated charge--discharge cycling reduces capacitance by as much as 50\% in a year~\cite{choi2022capos}. A smaller $C$ shortens the $RC$ time constant (\autoref{eq:discharge_time}), so the capacitor discharges faster; the same elapsed time yields a lower measured voltage, which the calibrated mapping reads as a longer duration---the clock overestimates progressively as it ages. This error has a fixed sign, so unlike measurement noise it does not average out but accumulates over power cycles, skewing the time estimate over the deployment lifetime (\autoref{fig:aging_expirations_cap_only}). Recalibration could correct it, but unattended deployments offer no such opportunity.

Both limitations are structural: measuring time with stored energy ties range to both energy and area, and ties reliability to a component that wears out. This motivates a timekeeping primitive built on a stable physical process, whose cost is independent of the measured interval and whose behavior does not drift with age.

\subsection{A Primer on MTJs}
\label{sec:mtj_primer}

\begin{figure}[t]
    \centering
    \begin{subfigure}{0.38\linewidth}
        \centering
        \includegraphics[width=0.95\linewidth]{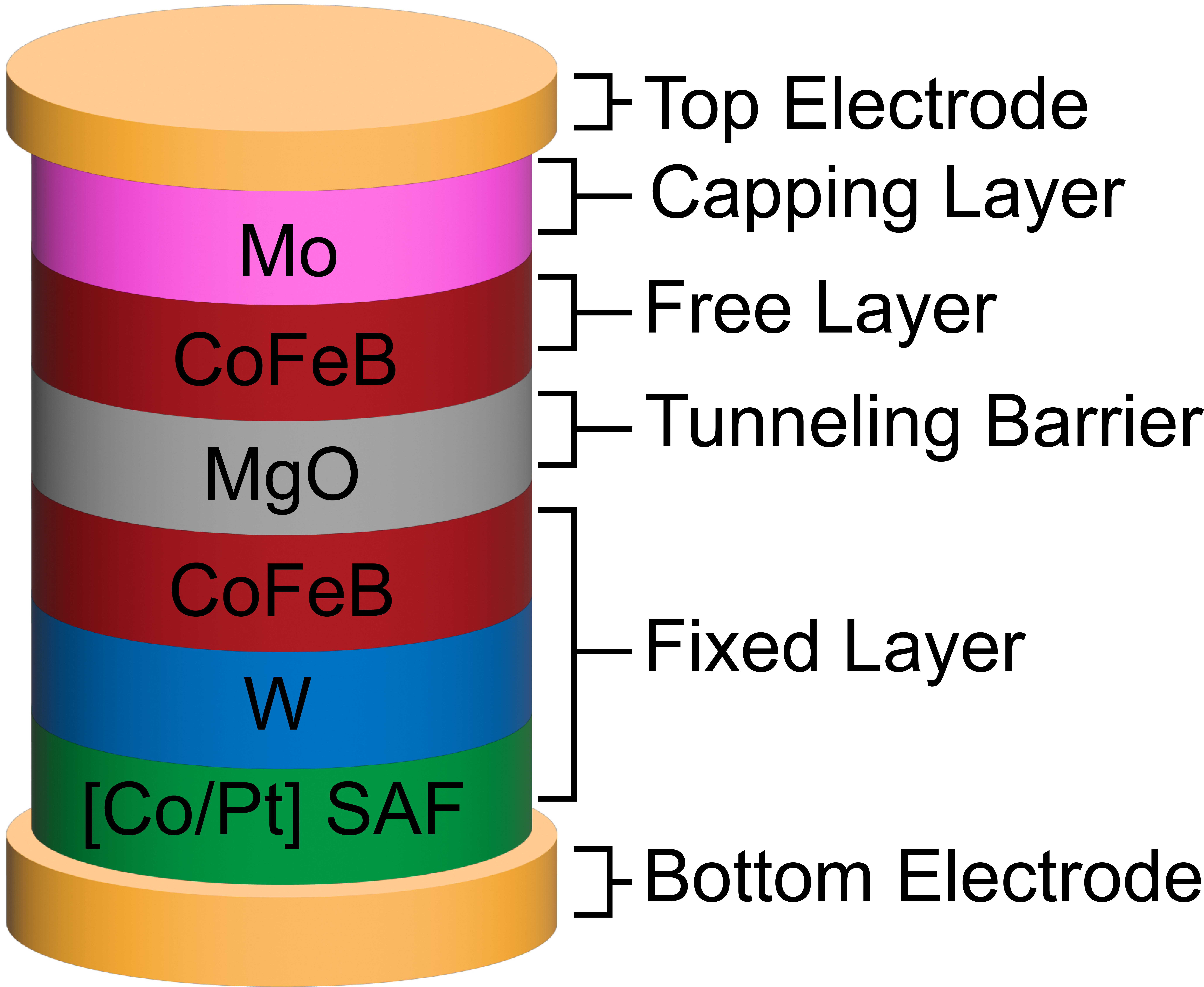}
        \caption{MTJ pillar schematic}
        \label{fig:mtj_stack}
    \end{subfigure}%
    ~
    \begin{subfigure}{0.62\linewidth}
        \centering
        \includegraphics[width=\linewidth]{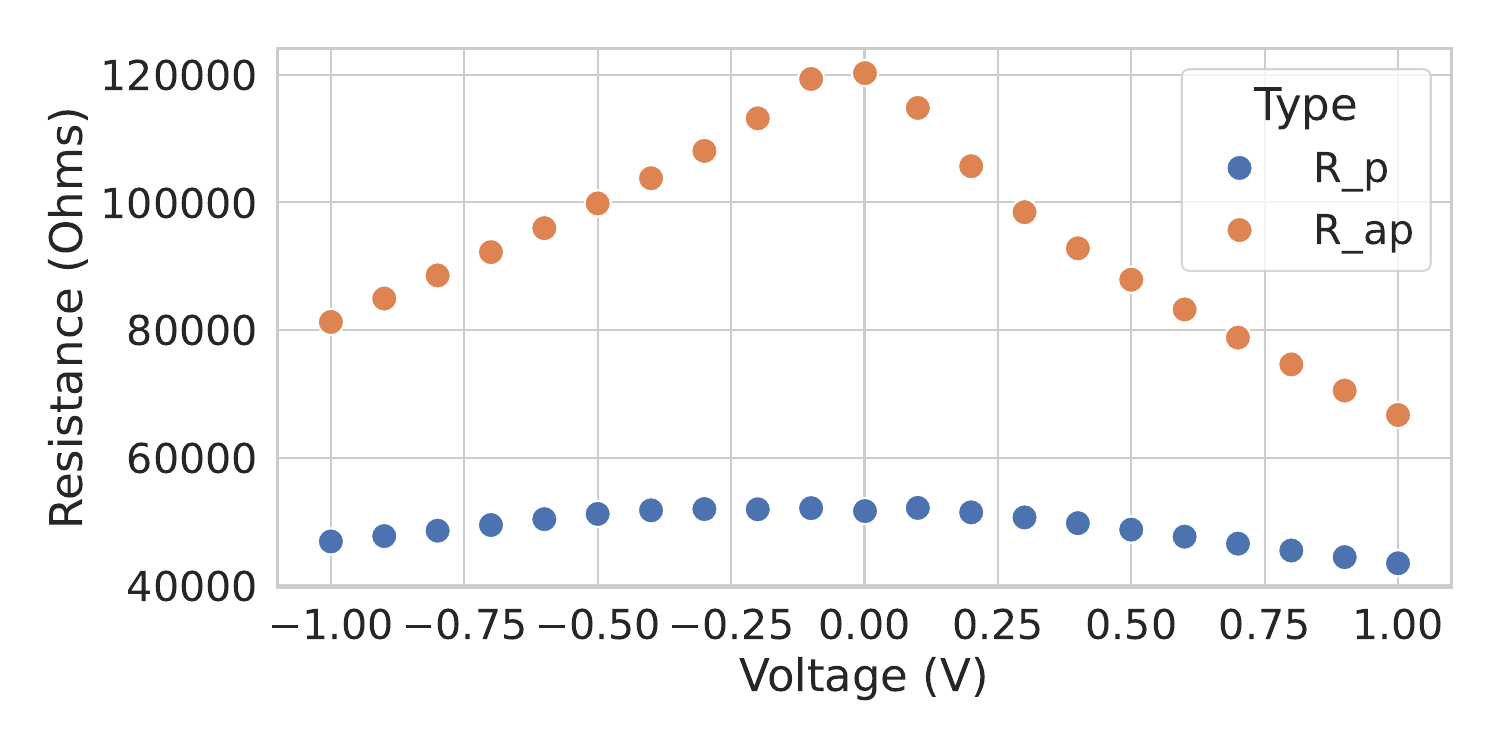}
        \caption{MTJ Resistance vs. Voltage}
        \label{fig:rp_rap}
    \end{subfigure}
    \caption{a) MTJ device stack. b) Resistance vs.\ applied voltage $V$ for P (\textcolor{Blue}{blue}) and AP (\textcolor{Orange}{orange}) states. Note drop in TMR with increasing $V$.}
    \Description{Left: cross-section diagram of magnetic tunnel junction with ferromagnetic layers and tunnel barrier. Right: resistance curve showing two distinct states with voltage dependence.}
    \label{fig:mtj_design_behav}
\end{figure}

A magnetic tunnel junction (MTJ) is a nanoscale spintronic device comprising two ferromagnetic layers separated by a thin insulating tunnel barrier (see \autoref{fig:mtj_stack}). The reference layer maintains a fixed magnetic orientation, relative to which the free layer occupies one of two stable magnetic alignment states: parallel (P) or antiparallel (AP). Due to quantum mechanical tunneling across the barrier, the junction's resistance depends strongly on this alignment---the P state exhibits lower resistance (R\textsubscript{P}) and the AP state exhibits higher resistance (R\textsubscript{AP}). This phenomenon is represented by the tunneling magnetoresistance ratio, $\textit{TMR} = (R_{AP} - R_P) / R_P$, which for the devices characterized in this work averages $125\%$ (see \autoref{fig:rp_rap}). 

The thermal stability of the free layer is governed by an energy barrier $E_b$, which determines the timescale for spontaneous state transitions. Switching behavior follows the N\'eel-Arrhenius relation, which states the mean dwell time ($\tau$) in a given state scales as
\begin{equation}\label{eq:neel-arrhenius}
    \tau = \tau_0 \cdot \exp\left(\nicefrac{E_b}{k_bT}\right)
\end{equation}
where $\tau_0$ is the attempt frequency, 
$k_b$ is the Boltzmann constant, and $T$ is temperature in Kelvin. By controlling $E_b$ during
fabrication, MTJs can be engineered with dwell times spanning many orders of magnitude. Devices with sufficiently low energy barriers switch stochastically under ambient thermal fluctuations alone, without requiring any applied write current \cite{liu2025application}. Crucially, $\tau$ is fixed by the device geometry and material stack, not by any stored charge---so the timescale neither costs energy to maintain nor drifts as the device is cycled, directly addressing the two failure modes above.

While individual MTJ switching events are inherently random, large arrays of identically configured MTJs exhibit well-characterized aggregate behavior. The energy barriers across devices in an array follow a near-normal distribution (\autoref{fig:dist_eb}), and as devices stochastically transition between states, the aggregate array imbalance ($|\textit{On} - \textit{Off}|$) and resistance decay in a smooth, reproducible, approximately exponential profile (\autoref{fig:single_decay} and \autoref{fig:multi_resistance}, respectively). By leveraging this statistical regularity, \sysname infers the elapsed time from a single resistance measurement of the array. 



\section{System Design}
\sysname (\autoref{fig:sys_overview}) is a batteryless timekeeping system that estimates off-time by leveraging the stochastic decay of MTJ arrays, whose resistance changes as devices transition between P and AP states. Each array is reset to a known state before power loss and decays passively during outages, requiring no active energy.

\subsection{\sysname Architecture and Data Flow}
\begin{figure*}
    \centering
    \includegraphics[height=2.5cm]{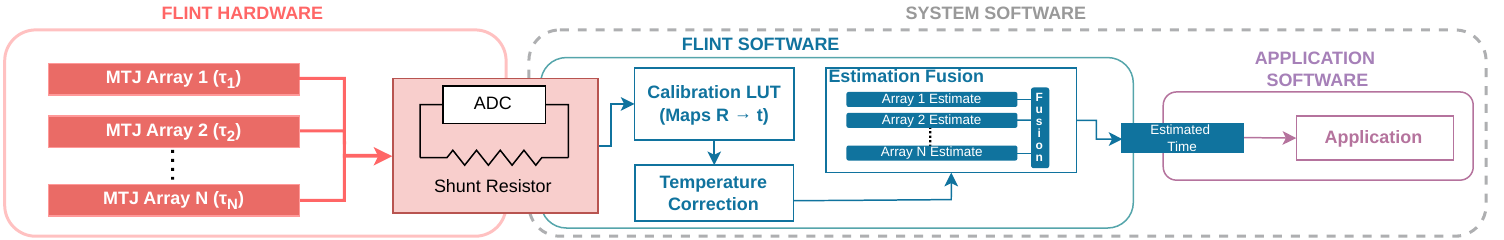}
    \caption{System architecture and dataflow. After the fusion stage, we obtain a single time estimate that is our result.}
    \Description{Block diagram showing FLINT data flow from multiple MTJ arrays through ADC readout, lookup table conversion, temperature correction, and fusion logic to produce time estimate.}
    \label{fig:sys_dataflow}
\end{figure*}

\autoref{fig:sys_dataflow} illustrates the architecture of \sysname. The system consists of multiple MTJ arrays, an ADC for resistance measurement, and a software layer running on the MCU. Each MTJ array is accessed through simple control circuitry that enables reset and resistance measurement. As shown in \autoref{fig:sys_overview}, a shutdown signal resets the array before power loss, while a measurement signal enables ADC-based resistance sampling upon power restoration. The measured resistance is mapped to elapsed time using a lookup table stored in memory, constructed during calibration. To improve robustness and extend the timekeeping range, \sysname employs multiple arrays with different dwell-time characteristics. Their outputs are combined using lightweight fusion logic to produce the final time estimate. Additional correction modules, such as temperature compensation, can be applied at runtime with minimal overhead, maintaining accuracy under varying environmental conditions (\textbf{R3}).

\subsection{Timekeeping Operation}

\begin{figure}
    \centering
    \begin{subfigure}[t]{0.48\linewidth}
        \centering
        \includegraphics[width=\linewidth]{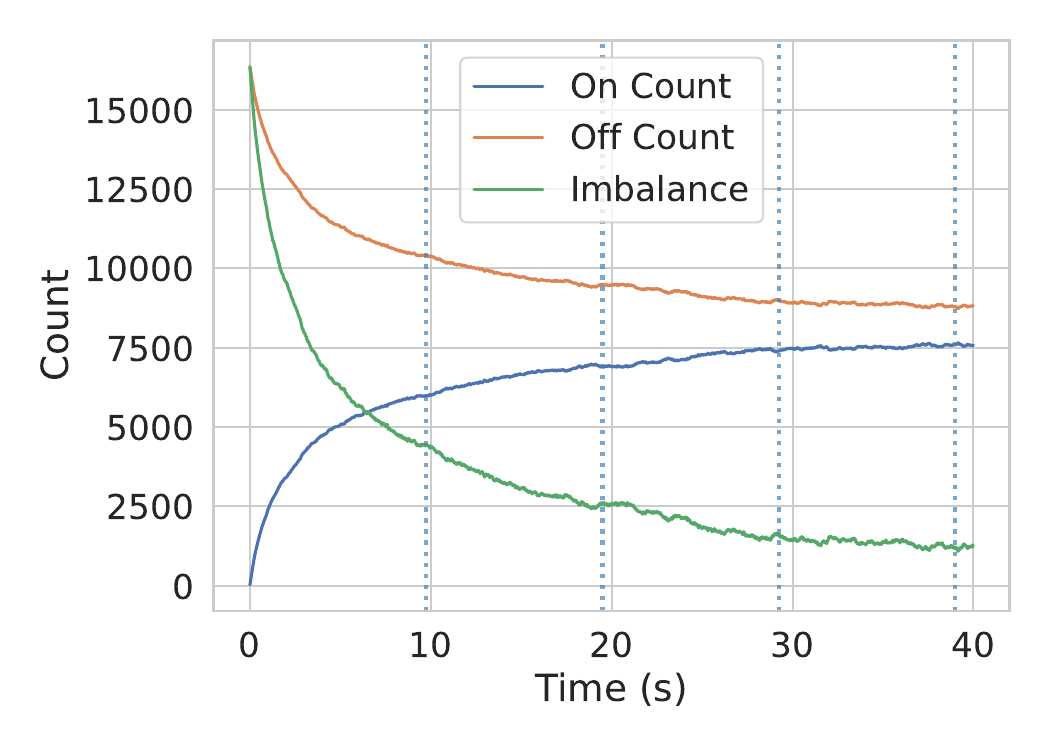}
        \caption{9.7\,s dwell time array. Green line is \textit{imbalance} = $|\textit{On} - \textit{Off}|$. Vertical dashed lines at dwell time multiples.}
        \label{fig:single_decay}
    \end{subfigure}
    \hfill
    \begin{subfigure}[t]{0.48\linewidth}
        \centering
        \includegraphics[width=\linewidth]{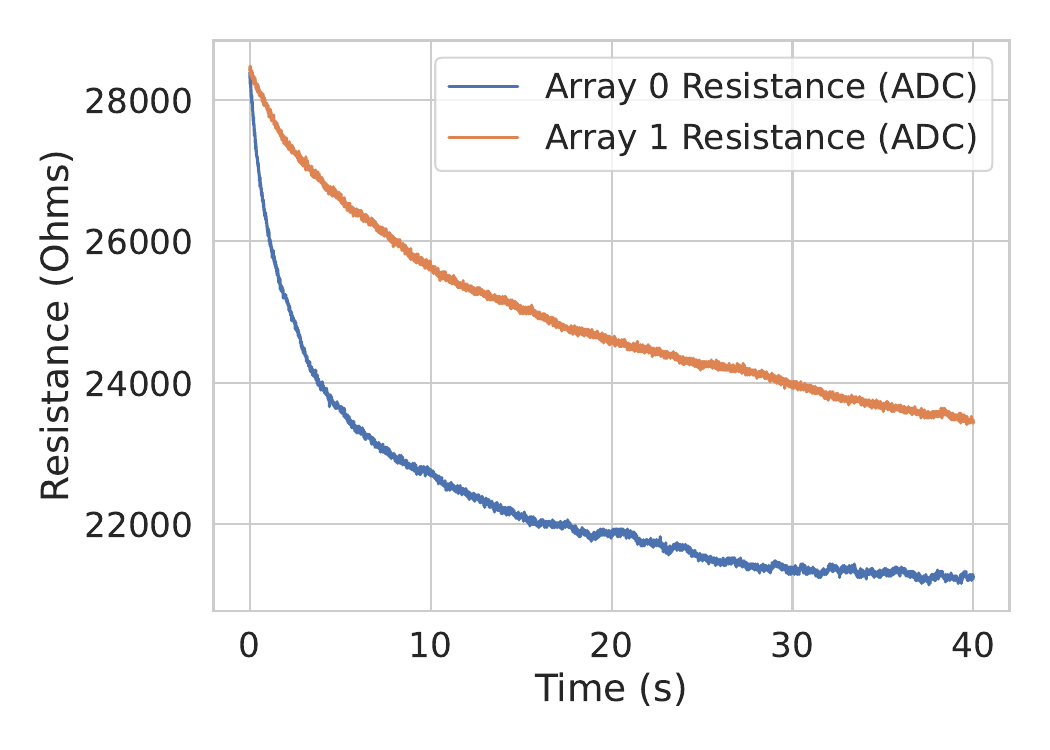}
        \caption{9.7-second (\textcolor{Blue}{blue}) and 72-second (\textcolor{Orange}{orange}) dwell time arrays. Starting resistance is \qty{28}{\kilo\ohm}.}
        \label{fig:multi_resistance}
    \end{subfigure}
    \caption{MTJ array decay curves ($64 \times 256$ arrays).}
    \Description{Left: imbalance decay over time showing exponential decay with periodic dwell time markers. Right: resistance decay for two different dwell time MTJ arrays showing monotonic change with time.}
    \label{fig:array_decays}
\end{figure}

\sysname estimates elapsed time across power failures through a sequence of operations spanning shutdown, off-time, and startup. Prior to power failure, each MTJ array is reset to a known initial state, where all devices are in the same magnetic state. During the off-time, the arrays evolve passively without energy consumption. Individual MTJs stochastically switch between P and AP states due to thermal effects, causing the imbalance between states to gradually decrease over time. As shown in \autoref{fig:array_decays}, this reduction in imbalance (\autoref{fig:single_decay}) drives a corresponding change in the overall array resistance (\autoref{fig:multi_resistance}), which evolves monotonically as the system approaches equilibrium. Different arrays exhibit different decay rates depending on their dwell-time characteristics, enabling the system to capture a wide range of time scales.

Upon power restoration, the system measures the resistance of each array using an ADC. This measured resistance is then mapped to elapsed time using a precomputed lookup table created in calibration, exploiting the monotonic relationship between resistance and time. Finally, when multiple arrays are used, their individual time estimates are combined using fusion logic to produce a single, robust estimate that balances resolution and timekeeping range.

\subsection{Design Choices}

\noindpar{Multi-array design.}
A single MTJ array provides limited timekeeping range due to saturation as devices approach equilibrium. To address this, \sysname employs multiple arrays with different dwell-time characteristics, enabling each array to capture a distinct portion of the time spectrum. Short dwell-time arrays provide fine resolution at small durations, while long dwell-time arrays extend the measurable range. This design allows \sysname to simultaneously achieve high resolution and long timekeeping range, which is difficult to realize with a single array.

\noindpar{Fusion across arrays.}
Combining estimates from multiple arrays is non-trivial due to differing stochastic dynamics and time sensitivities. Na\"ive averaging assumes a uniform and stationary behavior across arrays, which leads to large estimation errors. Instead, \sysname employs fusion mechanisms that prioritize arrays operating in their sensitive regions while discounting those that have saturated. This allows the system to dynamically balance short-term resolution and long-term accuracy, resulting in a more robust time estimate.

\noindpar{Lookup-based time inference.}
\sysname maps measured resistance to elapsed time using a precomputed lookup table derived during calibration. While analytical models could be used, the stochastic nature of MTJ switching and device variability makes closed-form modeling challenging and less robust. A lookup-based approach naturally captures these effects and enables efficient runtime estimation with minimal computational overhead.

\noindpar{Passive timekeeping.}
\sysname performs timekeeping entirely through passive device dynamics during off-time, without requiring active energy consumption. Unlike capacitor-based approaches that rely on charging dynamics and active measurement circuitry, \sysname leverages intrinsic stochastic switching behavior to encode time directly in device state. This enables zero energy overhead during outages and eliminates the need for dedicated timing hardware.

\noindpar{Range decoupled from energy.}
\begin{figure}
    \centering
    \includegraphics[width=0.9\linewidth]{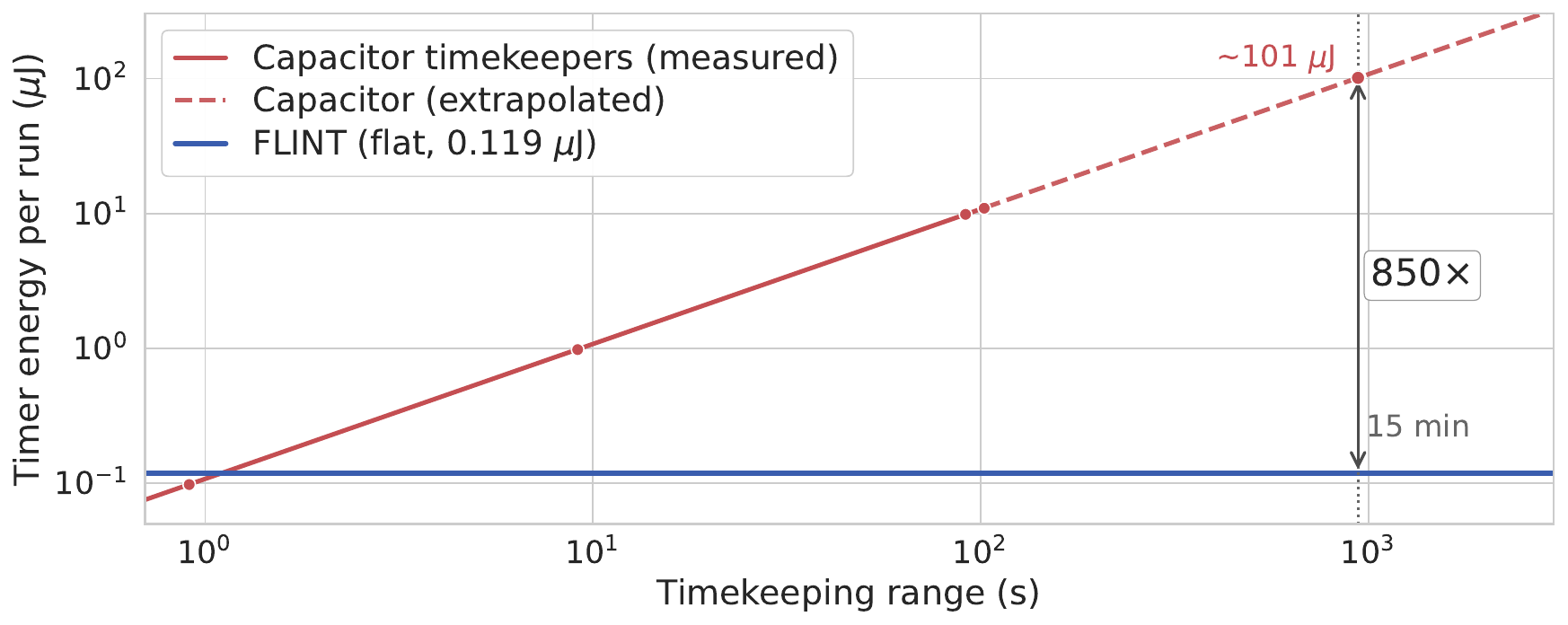}
    \caption{Timer energy vs.\ timekeeping range (log--log).}
    \Description{Log-log plot comparing timer energy consumption against timekeeping range for FLINT and capacitor-based approaches, showing linear and higher-order relationships.}
    \label{fig:energy_vs_range}
\end{figure}
A capacitor's range is set by how much charge it stores, so a longer range means a larger capacitor, more energy to charge it every cycle, and more board area---with energy scaling proportional to range (\autoref{fig:energy_vs_range}). \sysname breaks this coupling. Because elapsed time is derived from decay fixed by $E_b$, not stored charge, the energy to reset and measure an array is the same whether it tracks seconds or hours. Reaching a 15-minute range costs a capacitor clock roughly \qty{101}{\micro\joule} of timer energy and a bulky discrete part; \sysname pays \qty{0.119}{\micro\joule} at any range ($850\times$ less, see \autoref{sec:results-energy}). This is also what makes \sysname a ``one size fits all'' timekeeper: since an idle array draws no power and a $64\times256$ array occupies just \qty{655.36}{\um^2}, a chip can carry more arrays than any deployment needs, and the application selects which to use at runtime---trading range for resolution without changing hardware.

Together, these properties directly address \textbf{R1} (application-agnos\-tic operation): \sysname's energy and footprint are independent of the target interval---short gaps or multi-hour durations---so the same hardware serves any workload without re-sizing or re-configuration. Aging immunity satisfies \textbf{R2} (temporal stability): because $E_b$ is a device-geometry parameter unaffected by charge cycling, the decay timescale remains stable over the full deployment lifetime---no recalibration is needed as the device ages.

\section{Implementation}

\sysname consists of MTJs connected in a series-parallel configuration, surrounded by support hardware (\autoref{fig:mtj_array_circuitry}). Time is inferred from the measured resistance through a software pipeline, shown in \autoref{fig:sys_dataflow}. 


\subsection{Control Hardware}

\begin{figure}[htbp]
    \centering
    \includegraphics[width=0.9\linewidth]{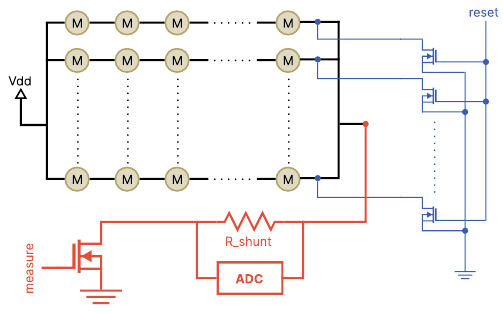}
    \caption{Supporting circuitry for a single array. Reset (\textcolor{boldblue}{blue}), measurement (\textcolor{newhorizon}{orange}). Width = series, height = parallel.}
    \Description{Schematic diagram showing MTJ array circuitry with reset control and measurement paths, including shunt resistor and ADC integration.}
    \label{fig:mtj_array_circuitry}
\end{figure}

We choose spin-transfer torque (STT) MTJs for \sysname because they are current-controlled and can be deterministically reset into a desired state, enabling reliable initialization. 
In addition, their $\tau$ can be controlled with adjustment of  $E_b$. 
STT devices are CMOS-compatible and energy-efficient, making them practical for scalable array fabrication and implementation~\cite{wang2013low}. The control hardware for a single MTJ array is depicted in \autoref{fig:mtj_array_circuitry}. It consists of the reset and measurement circuitry. The former \textbf{resets} and initializes the array just before shutdown, and is triggered when the system power reaches a predefined lower threshold. The latter is used to \textbf{measure} resistance of the array at startup, and is controlled by the SoC. These circuits are designed to minimize the required support hardware.

\textbf{Reset Circuitry:} A current pulse exceeding a pre-defined threshold (e.g. \qty{35}{\uA} for \qty{50}{\ns}) is used to reset the MTJ array to the AP state. To generate this pulse, we use an appropriately sized NMOS transistor for each row of the array (shown in \textcolor{boldblue}{blue} in \autoref{fig:mtj_array_circuitry}). The transistors are biased to operate in the saturation region and function as current sources, ensuring a consistent supply to each row. Since the MTJ resistance depends on the state and can differ across rows, each row requires a transistor to drive it.

\textbf{Measurement Circuitry:} To measure the array resistance, we determine the voltage across a series \textit{shunt resistor} ($R_{shunt}$), use it to compute the current through the shunt and MTJ array, and thus infer the array resistance. The shunt resistance is chosen to be approximately equal to the array resistance to maximize ADC dynamic range. One ADC channel is required per array. The system VDD is significantly lower than the reset voltage, limiting the measurement current. An NMOS transistor controls the circuit and is sized to minimize current through the array, avoiding unintended MTJ switching or degradation of TMR (see \autoref{fig:rp_rap}).

\subsection{Calibration}
Calibration is a vital process for ensuring high accuracy in varying deployment environments. To calibrate an array, it is first reset to its initial state. As the array decays, the resistance is sampled at a fixed interval, chosen as a small fraction of the dwell time (e.g., $\approx$0.05--0.1\%), to accurately capture the stochastic switching behavior of fast devices.
These measured values of resistance are associated with their respective elapsed time, and the process is repeated. Finally, the resistance measured across these runs is averaged at each timestep. This is used to build the lookup table that holds the information needed to infer time. Note the sampling process can be done in parallel for all arrays with a sufficiently fast ADC.

\begin{algorithm}
\small
\caption{Array Calibration Procedure}
\label{alg:calibration}
\begin{algorithmic}[1]
\Require Sampled data $\mathcal{D} = \{(t_1, R_1), \dots, (t_N, R_N)\}$, Table Size $M$
\State Sort $\mathcal{D}$ by time $t$, set $\mathbf{t} \gets [t_1,\dots, t_N]$ and $\mathbf{R}_{\text{raw}} \gets [R_1, \dots, R_N]$
\State $\mathbf{R}_{\text{iso}} \gets \text{IsotonicRegression}(\mathbf{t}, \mathbf{R}_{\text{raw}})$ \Comment{Filter noise, enforce monotonicity}
\State $P \gets \text{PCHIP}(\mathbf{t}, \mathbf{R}_{\text{iso}})$ \Comment{Construct continuous mapping}
\State $\mathbf{R}_{\text{LUT}} \gets \text{LinSpeace}(\min(P), \max(P), M)$
\For{$m = 1$ \textbf{to} $M$} \Comment{Invert PCHIP output to build calibration table}
    \State $\mathbf{t}_{\text{LUT}}[m] \gets \text{Solve}(P(t) = \mathbf{R}_{\text{LUT}}[m])$
\EndFor
\State \Return Calibration LUT $(\mathbf{R}_{\text{LUT}}, \mathbf{t}_{\text{LUT}})$
\end{algorithmic}
\end{algorithm}

Creating the lookup table (LUT) requires inverting the time vs. resistance data. 
We show our array calibration procedure in \autoref{alg:calibration}. Isotonic regression (L2)~\cite{chakravarti_IsotonicMedianRegression_1989, leeuw_IsotoneOptimizationPoolAdjacentViolators_2010, deleeuw_CorrectnessKruskalsAlgorithms_1977} is used to fit a monotonic step function on the collected data, enabling inversion. PCHIP interpolation~\cite{fritsch_MethodConstructingLocal_1984, moler_3Interpolation_2004} is applied to smooth the curve (L3), chosen as it preserves monotonicity. Finally, the LUT is built by creating a linear space of resistance values and sampling the PCHIP curve with them (L4-6). 

The closest resistance value in the lookup table is found using a binary search. Linear interpolation is then used to compute the resulting time estimate. If the resistance value falls outside the table, it is clamped to the closest boundary. 

\subsection{Temperature Correction}
\label{sec:temp-correction}


To correct for temperature variations, we use the N\'eel-Arr\-henius equation to scale our estimates based on the operating temperature $T_{op}$. Given the design energy barrier value $\widehat{E_b}$ (as a temperature-dependent dimensionless quantity, rather than units of $kT$) and calibration temperature $T_{cal}$, the corrected time estimate $t_{corr}$ is computed from the raw time estimate $t_{raw}$ as follows:
\begin{equation}
    t_{corr} = t_{raw} \cdot \exp\left(\widehat{E_b} \cdot \left(\nicefrac{1}{T_{\text{op}}} - \nicefrac{1}{T_{\text{cal}}}\right)\right)
\end{equation}

\subsection{Estimate Fusion}\label{sec:impl-fusion}

\begin{algorithm}
\caption{Shortest Active Fusion}
\label{alg:fusion-shortest-active}
\small
\begin{algorithmic}[1]
\Require Estimates $T = [t_1, \dots, t_N]$, Dwell times $D = [\tau_1, \dots, \tau_N]$
\State $G \gets 1.25$ \Comment{Reliability gate multiplier}
\State $\tau_{\text{min}} \gets \infty$, $idx_{\text{best}} \gets \arg\max_i D[i]$ \Comment{Fallback to longest dwell}
\For{$i = 1$ \textbf{to} $N$}
    \State \textbf{update} $idx_{\text{best}} \gets i$, $\tau_{\text{min}} \gets \tau_i$ \textbf{if} $t_i < G \cdot \tau_i$ \textbf{and} $\tau_i < \tau_{\text{min}}$
\EndFor
\State \Return $t_{idx_{\text{best}}}$
\end{algorithmic}
\end{algorithm}

Combining time estimates from each array must account for the variation in range and accuracy inherent to each array. A simple average of the estimated time does not capture these differences. 
To resolve this, we implement two fusion algorithms. The first chooses the shortest dwell time array that is still valid (i.e., the imbalance is still high), as shown in  \autoref{alg:fusion-shortest-active}. This is based on the fact that arrays with shorter dwell times can measure at a higher resolution and maximize accuracy. To determine if an array is valid, we compare its estimate $t_{est}$ to its dwell time $\tau$: if $t_{est} < G \cdot \tau$, the array is considered valid. The gate factor $G = 1.25$ identifies the optimal range for each array.
If none of the arrays are deemed valid, the estimate from the array with the longest dwell time is used. 

\begin{algorithm}
\caption{Linear-Decay Fusion (Linear \& Sqrt)}
\label{alg:fusion-decay}
\small
\begin{algorithmic}[1]
\Require Estimates $T = [t_1, \dots, t_N]$, Decay cutoff factor $C = 2$, \\Dwell times $D = [\tau_1, \dots, \tau_N]$, Mode $M \in \{\text{Linear}, \text{Sqrt}\}$
\State $W_{\text{total}} \gets 0, \ S_{\text{w}} \gets 0, \textit{idx}_{\text{longest}} \gets \arg\max_i D[i]$
\For{$i = 1$ \textbf{to} $N$}
    \State $s_i \gets \tau_i$ \textbf{if} $M = \text{Linear}$ \textbf{else} $\sqrt{\tau_i}$ \Comment{Scale factor} 
    \State $gate_i \gets \max\left(0, 1 - \nicefrac{t_i}{(C \cdot s_i)}\right)$, $w_i \gets \nicefrac{gate_i}{s_i}$
    \State $W_{\text{total}} \gets W_{\text{total}} + w_i$,  $S_{\text{w}} \gets S_{\text{w}} + w_i \cdot t_i$
\EndFor
\State \Return $t_{idx_{\text{longest}}}$ \textbf{if} $W_{\text{total}} = 0$ \textbf{else}  $S_{\text{w}} / W_{\text{total}}$ \Comment{Fallback to longest.}
\end{algorithmic}
\end{algorithm}

The second algorithm performs a weighted average, as shown in \autoref{alg:fusion-decay}. We bias these towards the shorter-range arrays to maximize accuracy. We formulate a linear weight curve for each array with y-intercept $\nicefrac{1}{s_i}$ and x-intercept  $C\cdot s_i$. 
We set $C$ to 2 as arrays are known to be mostly effective up to 2\texttimes{} their dwell time, and compute $s_i$ proportional to the array dwell time $\tau_i$. 
We define two operational modes, linear and sqrt, setting $s_i = \tau_i$ and $s_i = \sqrt{\tau_i}$, respectively. Sqrt mode assigns a higher weight to short-range arrays at the start. The weight for each estimate $t_i$ is calculated from the linear weight curve; the weighted average determines the final time estimate.
The longest dwell-time array is the fallback if all weights are zero.


\section{Evaluation Methodology}

We evaluate our design through a combination of real-world experiments and simulation. Our simulations were informed and validated by the real-world data we gathered.

\subsection{Real-World MTJ Experiments}

To confirm that physical MTJs exhibit the stochastic characteristics required for \sysname's timekeeping approach, we characterize 21 devices fabricated using the same stack composition and layer thicknesses as reported in \cite{athas2025statistical}. The measured devices are voltage-controlled MTJs (V-MTJs), in which the energy barrier ($E_b$) can be tuned via an applied DC voltage through the voltage-controlled magnetic anisotropy (VCMA) effect. This allows a single set of physical devices to be characterized across multiple target dwell time regimes without requiring separate fabrication runs, emulating the behavior of different fixed-$E_b$ devices. Although the final system operates using STT for switching and reset, VCMA is used here purely as a characterization tool to modulate $E_b$; it does not alter the underlying thermal switching dynamics measured.

We configured each device for two operating points, $\tau$ = 10s and 25 ms, and recorded the time between switching events  (10 s is generally useful; 25 ms targets high resolution). From the measured switching event sequence, we computed per-device $\tau$ and applied the N\'eel-Arrhenius relation to extract the corresponding $E_b$. In practice, MTJs with $\tau$ spanning milliseconds to hundreds of seconds can be created by adjusting the free layer volume, anisotropy, or material stack during fabrication \cite{li2015thermally}. This corresponds to $E_b$ on the order of 10-40$k_bT$,
making our target operating points representative of deployable \sysname arrays.

\subsection{Array Simulations}\label{sec:eval-arr-sims}

We evaluate \sysname hierarchically at the array level. We first study a single MTJ array to characterize fundamental accuracy--range trade-offs and sensitivity to design parameters. We then extend the analysis to multi-array configurations, where heterogeneous arrays are composed to improve timekeeping range and robustness.

\begin{figure}[t]
    \centering
    \includegraphics[width=0.9\linewidth]{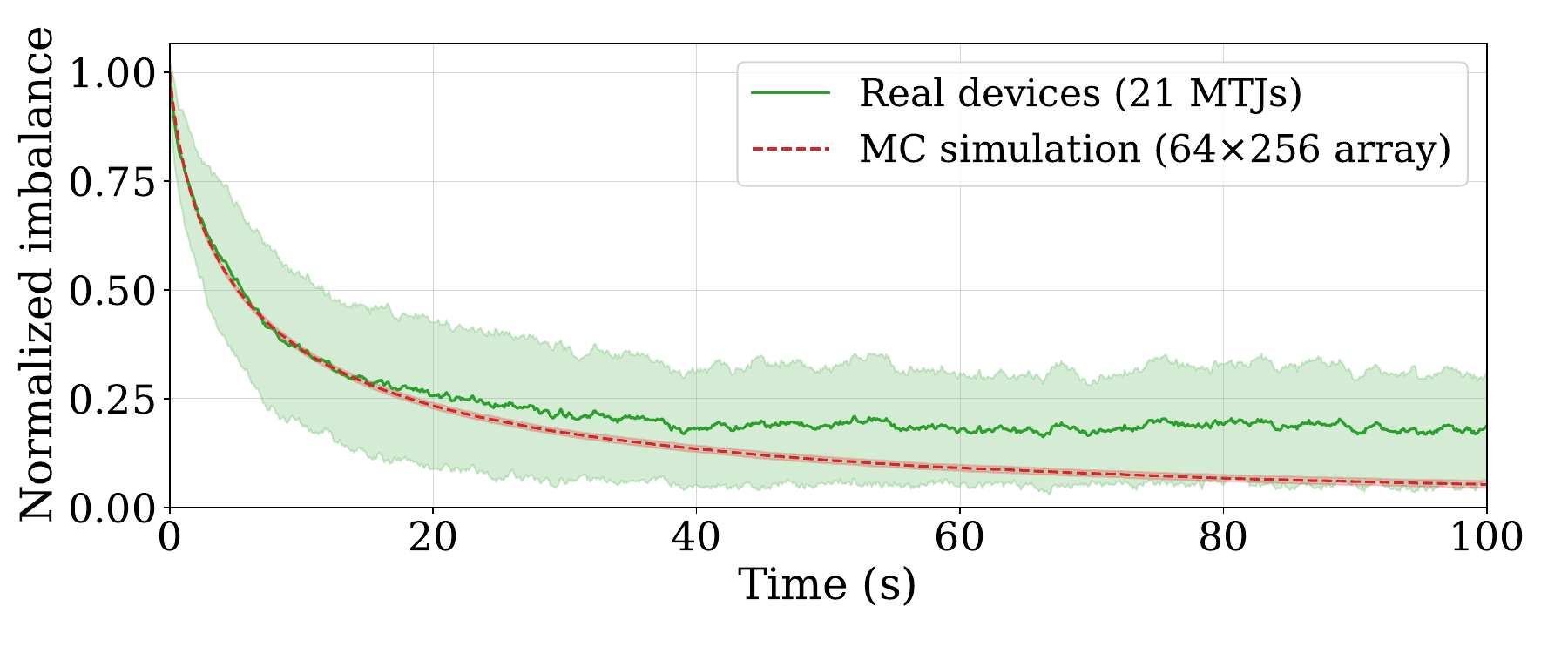}
    \caption{Sim validation: normalized imbalance for 21 fabricated MTJs (\textcolor{Green}{green}, $\pm1\sigma$) vs.\ Monte Carlo sim of $64\times256$ array (\textcolor{Red}{red} dashed).}
    \Description{Graph comparing normalized MTJ array imbalance decay between measured hardware (green band with error bars) and simulator predictions (red dashed line).}
    \label{fig:hw_validation}
\end{figure}

We model FLINT using Monte Carlo simulations that capture the stochastic switching behavior of MTJs and their aggregate effect on time estimation. Device characteristics are derived from validated hardware models~\cite{garcia-redondo_CompactModelScalable_2021, garcia-redondo_FokkerPlanckSolverModel_2021} and  experimental measurements (e.g., $E_b$, TMR). \autoref{fig:hw_validation} confirms that the simulator accurately reproduces measured device behavior (validation methodology in \autoref{sec:app-validation}). 

We perform sensitivity analyses over key design parameters, sweeping each parameter independently and reporting results as geometric means over 100 trials to ensure statistical robustness. Since ADC sampling latency ($\approx \qty{1}{\us}$~\cite{texasinstruments_MSP430FR596xMSP430FR594xMixedSignal_2018}) is negligible relative to MTJ dwell times, it is omitted.

\begin{table}[htbp]
    \centering
    \footnotesize
    \caption{Device and System Configurations.}
    \begin{tabular}{cc}
        \toprule
         \textbf{Parameter} & \textbf{Default Value} \\  
         \midrule
         \rowcolor{black!3}
         ADC Bits & 12\\
         ADC Noise & 1 ADC Step\\
         \rowcolor{black!3}
         MTJ Array Size & $64 \times 256$ \\
         Calibration Run Count & 20\\
         \rowcolor{black!3}
         Calibration Table Size & 128 entries/array\\
         $E_b$ std. dev. & 6.3\% \\
         \rowcolor{black!3}
         Temperature & \qty{293}{\K}\\
         $V_{meas}$ & \qty{0.052}{\V} per MTJ (VDD = \qty{3.3}{\V})\\
         \rowcolor{black!3}
         TMR & 124.49\%, $\sigma=7.81\%$\\
         $R_{P}$ & \qty{52.66}{\kilo\ohm}, $\sigma=\qty{3.73}{\kilo\ohm}$\\
         \rowcolor{black!3}
         $R_{shunt}$ & \qty{27.5}{\kilo\ohm}\\
         Fusion Method & LinearSqrtDecay\\
         \bottomrule
    \end{tabular}
    \label{tab:device-configs}
\end{table}

\begin{figure*}[htbp]
    \centering
    \begin{subfigure}{0.33\linewidth}
        \centering
        \includegraphics[width=\linewidth]{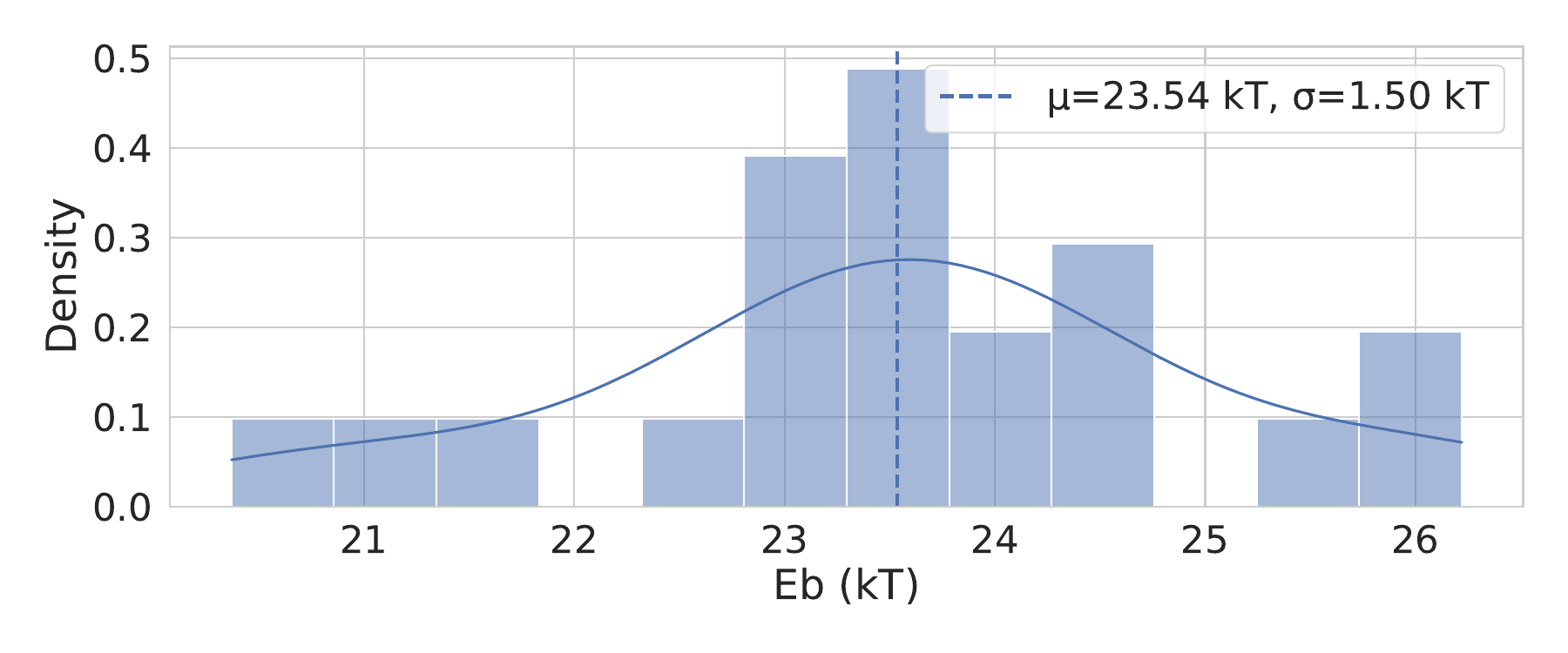}
        \caption{Energy barrier (units of $kT$).}
        \label{fig:dist_eb}
    \end{subfigure}
    ~
    \begin{subfigure}{0.33\linewidth}
        \centering
        \includegraphics[width=\linewidth]{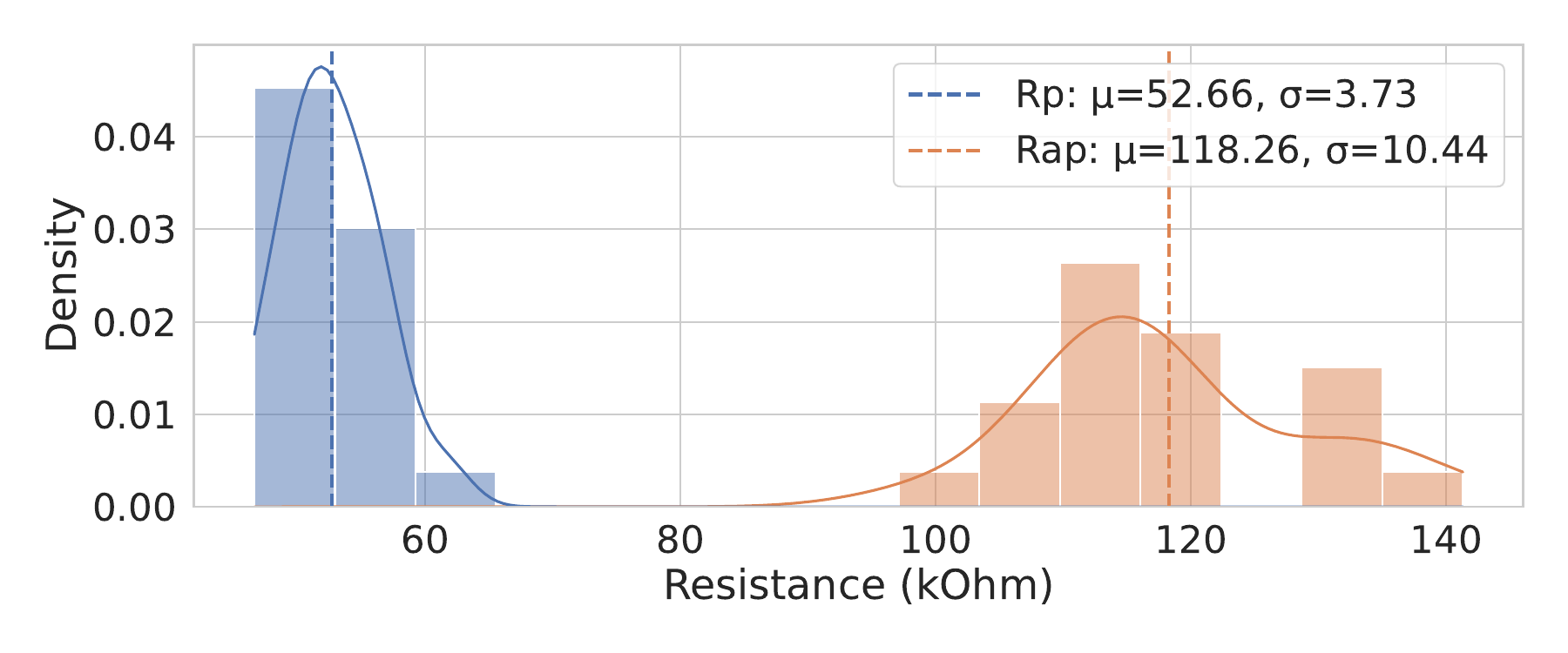}
        \caption{On ($R_P$) and off ($R_{AP}$) resistances.}
        \label{fig:dist_resist}
    \end{subfigure}
    ~
    \begin{subfigure}{0.33\linewidth}
        \centering
        \includegraphics[width=\linewidth]{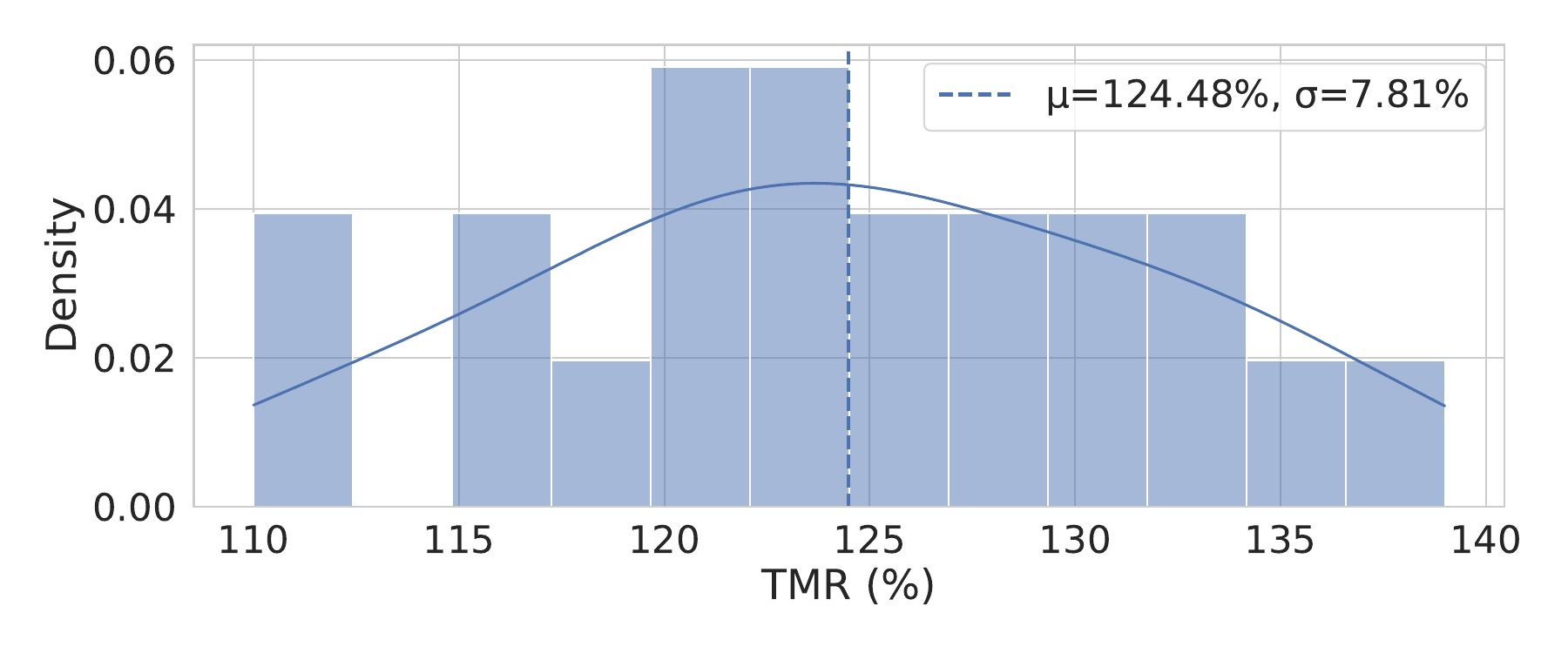}
        \caption{TMR ratio.}
        \label{fig:dist_tmr}
    \end{subfigure}
    \caption{MTJ parameter distributions for devices with a \qty{10}{\s} dwell time.}
    \Description{Three histogram plots showing device-to-device variability in energy barrier, resistance values (parallel and antiparallel states), and tunneling magnetoresistance ratio.}
    \label{fig:mtj_dists}
\end{figure*}

\textbf{Single-Array Experiments.}
To quantify energy consumption and guide the design of our control hardware, we use SPECTRE-based hardware simulations~\cite{garcia-redondo_CompactModelScalable_2021, garcia-redondo_FokkerPlanckSolverModel_2021}. 
We then evaluate FLINT using sensitivity analyses across key design dimensions, including sensing parameters (ADC precision, noise, and measurement voltage), array configuration (size and aspect ratio), calibration parameters (run count and table size), device variability ($E_b$ variation), and environmental factors (temperature). Default values are listed in \autoref{tab:device-configs}, and we set $E_b = 23\,kT$ ($\tau = \qty{9.74}{\s}$). Unless otherwise stated, parameters are chosen based on hardware measurements (e.g., $E_b$, TMR, $R_{P}$) or representative MCU capabilities (e.g., 12-bit ADC~\cite{texasinstruments_MSP430FR596xMSP430FR594xMixedSignal_2018}). Detailed impacts of these parameters are discussed in \autoref{sec:results}.


\textbf{Multiple-Array Experiments.}
We evaluate a multi-array configuration consisting of four MTJ arrays with  heterogeneous energy barriers of $23/24/25/26\,kT$ ($\tau = 9.74/26.5/72.0/196\,\text{s}$) and identical default parameters (\autoref{tab:device-configs}). We focus on two key factors: robustness to uncalibrated temperature variation and the effectiveness of different fusion methods. We compare our proposed fusion strategies (\autoref{sec:impl-fusion}) against a simple averaging baseline representative of lightweight implementations.
We also attempted to add HARC-Lite~\cite{deep_HARCHeterogeneousArray_2020}, but could not reproduce the results using the pseudocode.



\textbf{Metrics.}
We report several metrics: (i) percent error over time, (ii) absolute error due to $E_b$ variation for short durations ($[0,5]$\,s), (iii) time to 10\% error, the effective timekeeping range, (iv) energy overhead, and (v) latency.


\subsection{Comparison with State-of-the-Art}\label{sec:eval-sota}
We compare \sysname against prior timekeeping approaches using two complementary evaluations. First, we analyze system-level characteristics, including measurement range, energy consumption, startup latency, and resolution. Second, we evaluate time-estimation accuracy using four real energy-harvesting traces from kinetic and solar sources~\cite{geissdoerfer2022learning} to capture realistic operating conditions.

For system-level comparison, we use three capacitors of \qty{22}{\nF}, \qty{220}{\nF}, and \qty{2.2}{\uF} for CUSTARD~\cite{hester_PersistentClocksBatteryless_2016}, HARC~\cite{deep_HARCHeterogeneousArray_2020}, and CHRT~\cite{dewinkel_ReliableTimekeepingIntermittent_2020}. We assume 0805 capacitors to estimate the size. We configure three arrays in \sysname with dwell times to achieve the same range as the other systems. The calibration table contains 1024 entries to ensure comparability with CHRT~\cite{dewinkel_ReliableTimekeepingIntermittent_2020}. All \sysname results are aggregated across 100 runs. For trace-based evaluation of CHRT and \sysname, we use four real energy-harvesting traces and evaluate performance under different energy buffer sizes of \qty{10}{\uF}, \qty{100}{\uF}, and \qty{1}{\mF}, reflecting a range of intermittent operating conditions.

\subsection{Long-Term Aging Comparison}\label{sec:eval-aging}
We compare \sysname against an ideal capacitor-based timekeeper based on the C3 (\qty{2.2}{\uF}) configuration from \autoref{sec:eval-sota}, which provides the longest range (\qty{91.3}{\s}) among the evaluated capacitor-based designs. We model capacitance degradation (following ~\cite{choi2022capos}) as $C(d)=C_0(1-\nicefrac{\alpha d}{365})$, where $d$ is the deployment day and $\alpha$ is the annual degradation rate. We evaluate conservative ($\alpha = 0.20$, \qty{20}{\percent}/yr) and worst-case ($\alpha = 0.50$, \qty{50}{\percent}/yr) aging scenarios. To isolate the effect of aging, we attribute all capacitor clock error solely to capacitance degradation and exclude measurement, quantization, and leakage effects, intentionally favoring the capacitor-based approach. As capacitance decreases, the system range scales with $C(d)$, and measured time is overestimated by $\nicefrac{C_0}{C(d)}$. In contrast, \sysname's MTJ energy barriers ($E_b$) are unaffected by charge-discharge cycling, and its timekeeping behavior therefore remains unchanged throughout the deployment. To quantify application-level impact, we use the \sysname configuration from \autoref{sec:eval-sota} and evaluate a representative sensing task scheduled every \qty{60}{\s} (a common sensing interval), counting re-sample mistakes over a simulated day.

\section{Results}\label{sec:results}

We evaluate \sysname through array-level simulations, energy analysis, comparison with SOTA, aging analysis, and case studies. Results are presented in the following sections.



\subsection{Device Characterization}

The resistance and TMR characteristics of the devices are shown in \autoref{fig:mtj_dists}. The R\textsubscript{P} and R\textsubscript{AP} distributions (\autoref{fig:dist_resist}) confirm well-separated resistance states with low device-to-device variability, and a mean TMR of $124.45\%$ ($\sigma = 7.81\%$) (\autoref{fig:dist_tmr}) provides sufficient contrast for reliable ADC-based resistance measurement.

The extracted $E_b$ distribution at both operating points is shown in \autoref{fig:dist_eb} and \autoref{fig:dist_eb_25ms}. At the \qty{10}{\s} target, devices have a mean $E_b = 23.5\,kT$ ($\sigma = 1.50$); at \qty{25}{\ms}, the mean shifts to $E_b = 17.0\,kT$ ($\sigma = 1.17$). Both distributions are well-described by a normal centered near the target $E_b$, confirming that device-to-device variation is bounded and predictable across dwell-time regimes. These measurements are the basis for the simulation parameters in \autoref{sec:eval-arr-sims}.

\subsection{Single Array Trade-off Analysis}
\label{sec:results-single-array}

We study the impact of key design parameters on timekeeping accuracy and identify configurations that achieve $\ge \qty{40}{\s}$ of timekeeping within 10\% mean error. Further analysis is also provided in \autoref{sec:app-more-sims}.

 

\begin{figure}[t]
    \centering
    \includegraphics[width=\linewidth]{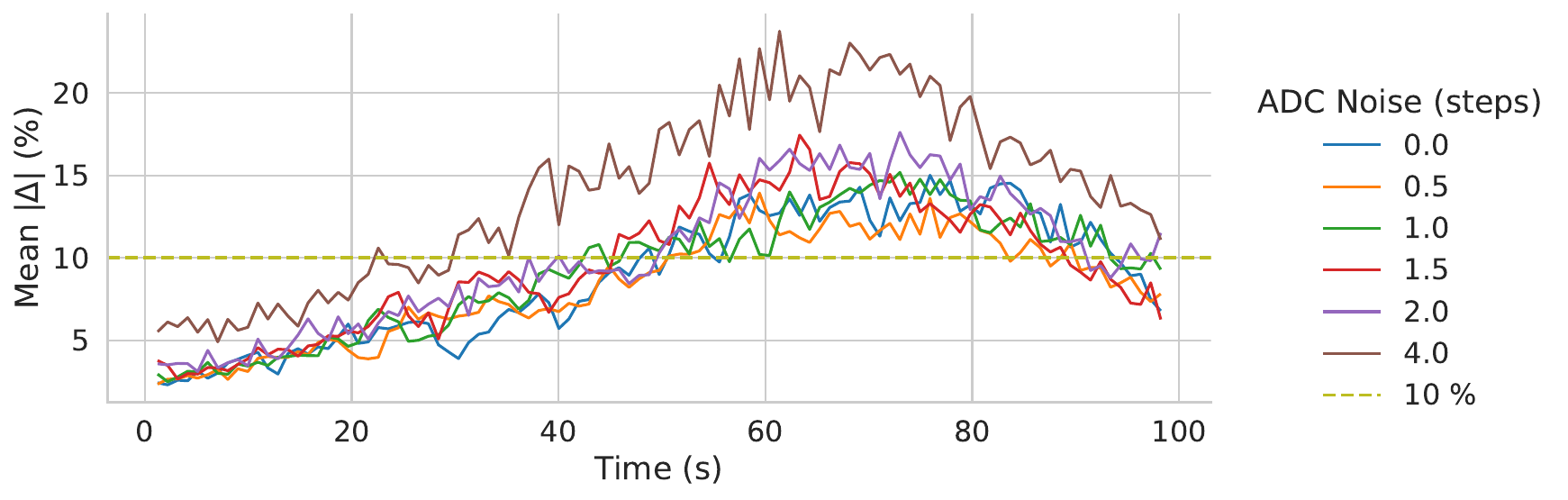}
    \Description{Line plot showing estimation error percentage versus elapsed time. Multiple colored lines represent different ADC noise levels from 0 to 5 millivolts. System maintains low error at all times up to 2.4 mV noise, then error increases at higher noise levels.}
    \caption{Timekeeper error with varying ADC noise.}
    \label{fig:eb23_adc_noise}
\end{figure}

\noindpar{Impact of ADC Noise.} We evaluate the impact of ADC noise on timekeeping accuracy, as shown in \autoref{fig:eb23_adc_noise}. Noise is injected at the MTJ output in multiples of the ADC quantization step, as only noise exceeding this resolution affects the measured voltage. We observe that the system remains robust up to 2.4mV (3 quantization steps on a 12-bit ADC at 3.3V VDD), beyond which the timekeeping range degrades by nearly 50\%. This is explained by the intrinsic separation between MTJ states: even under worst-case variation, the $R_P$–$R_{AP}$ voltage difference remains 6× larger (see \autoref{fig:dist_resist}) than the ADC quantization step, providing sufficient noise margin.

\begin{figure}[t]
    \centering
    \includegraphics[width=\linewidth]{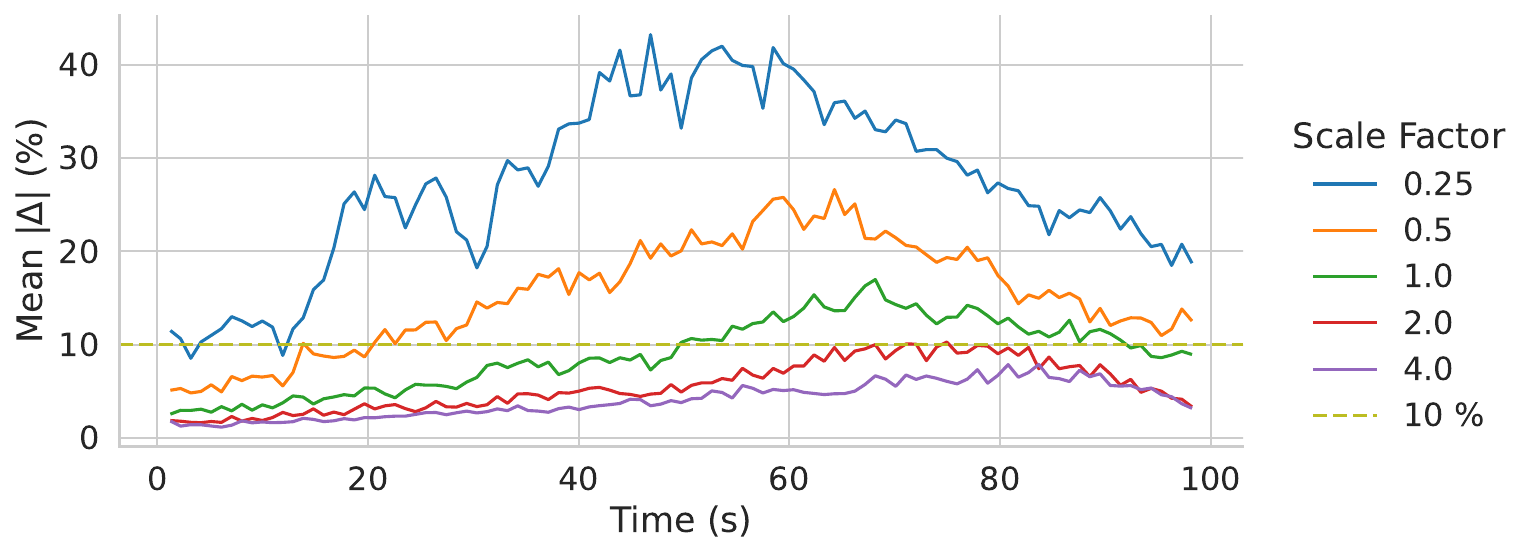}
    \Description{Line plot showing estimation error percentage versus elapsed time. Multiple colored lines represent array scaling factors from 0.25 to 4 times the base 64 by 64 size. Error decreases with larger arrays.}
    \caption{Timekeeper error with varying array scales.}
    \label{fig:eb23_array_scale}
\end{figure}

\noindpar{Impact of Array Size.} In \autoref{fig:eb23_array_scale}, we evaluate the impact of array size by scaling a 64×64 MTJ array. Increasing the number of MTJs improves precision and timekeeping accuracy. However, larger arrays require higher reset voltages: for example, a 64×64 array may require $\sim$10\,V per row, while a 256×256 array would increase this requirement to $\sim$40\,V, which may be impractical for intermittent battery-free systems. This highlights a fundamental trade-off: array size can be increased to improve accuracy, but is ultimately constrained by the reset voltage supported by the system.

\begin{figure}[t]
    \centering
    \includegraphics[width=\linewidth]{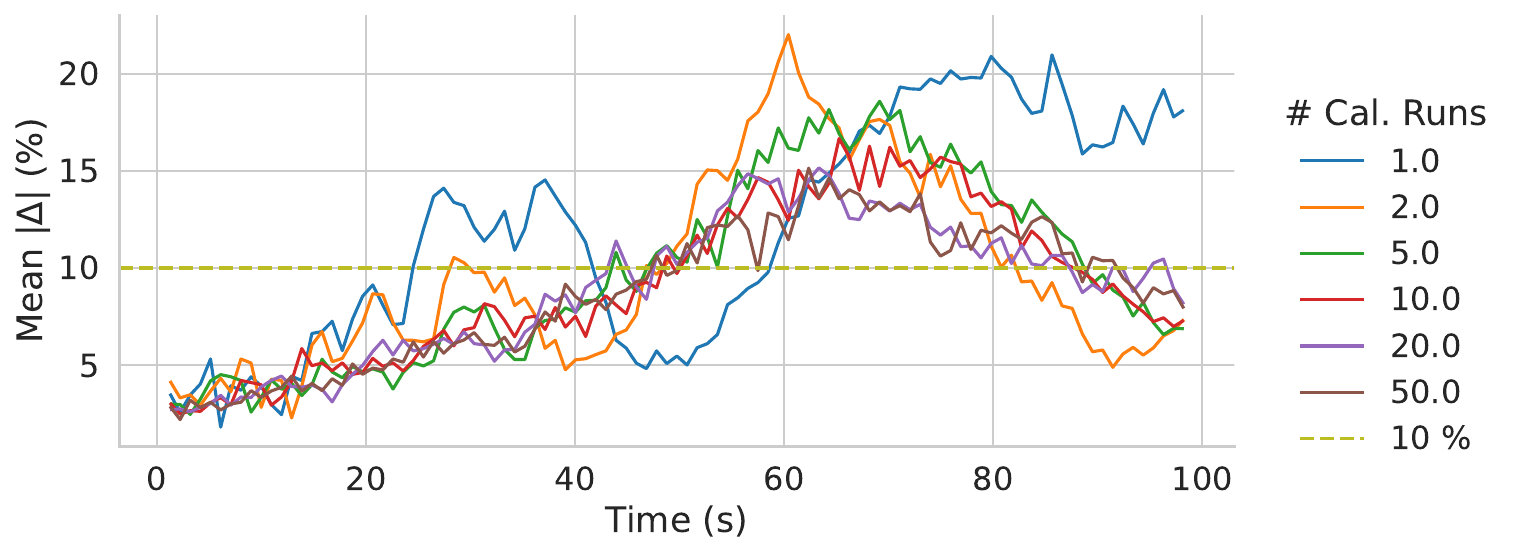}
    \Description{Line plot showing estimation error percentage versus elapsed time. Multiple colored lines represent different numbers of calibration runs from 1 to 20. Error decreases significantly between 1 and 10 runs, then plateaus.}
    \caption{Timekeeper error with varying calibration runs.}
    \label{fig:eb23_cal_runs}
\end{figure}

\begin{figure}[t]
    \centering
    \includegraphics[width=\linewidth]{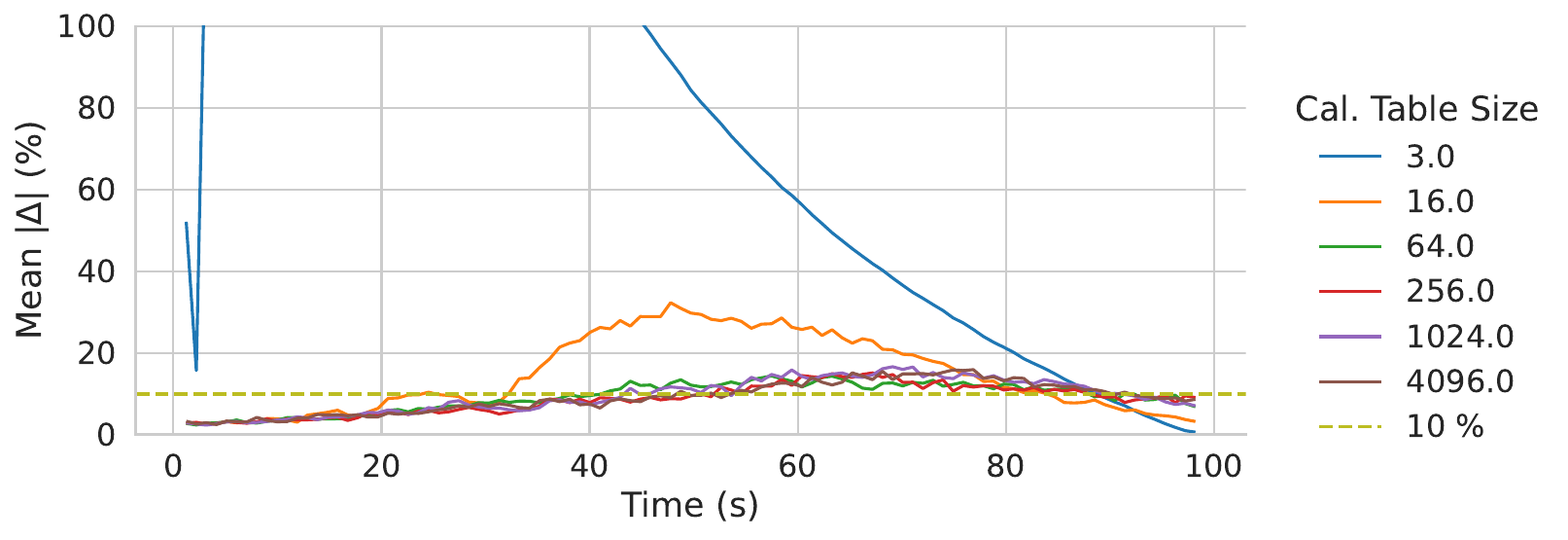}
    \Description{Line plot showing estimation error percentage versus elapsed time. Multiple colored lines represent calibration table sizes ranging from 16 to 256 entries. Error improves with larger tables but shows diminishing returns beyond 64 entries.}
    \caption{Impact of varying calibration table size.}
    \label{fig:eb23_cal_table_size}
\end{figure}

\noindpar{Impact of Calibration.} We evaluate the impact of calibration effort by varying both the number of calibration runs and the size of the calibration table, as shown in \autoref{fig:eb23_cal_runs} and \autoref{fig:eb23_cal_table_size}, respectively. Increasing calibration runs reduces noise in the learned mapping, while larger tables improve resolution through finer-grained resistance-to-time mapping. We observe diminishing returns beyond moderate configurations: 5–10 calibration runs are sufficient, and at least 64 table entries are required to achieve a \qty{40}{\s} timekeeping range within 10\% error. We select a 128-entry table (\qty{1}{\kibi\byte}) as our default. While increasing these parameters improves accuracy, it also introduces overhead in calibration time, memory, and latency, similar to prior approaches~\cite{dewinkel_ReliableTimekeepingIntermittent_2020,deep_HARCHeterogeneousArray_2020}.

\begin{figure}[t]
    \centering
    \includegraphics[width=\linewidth]{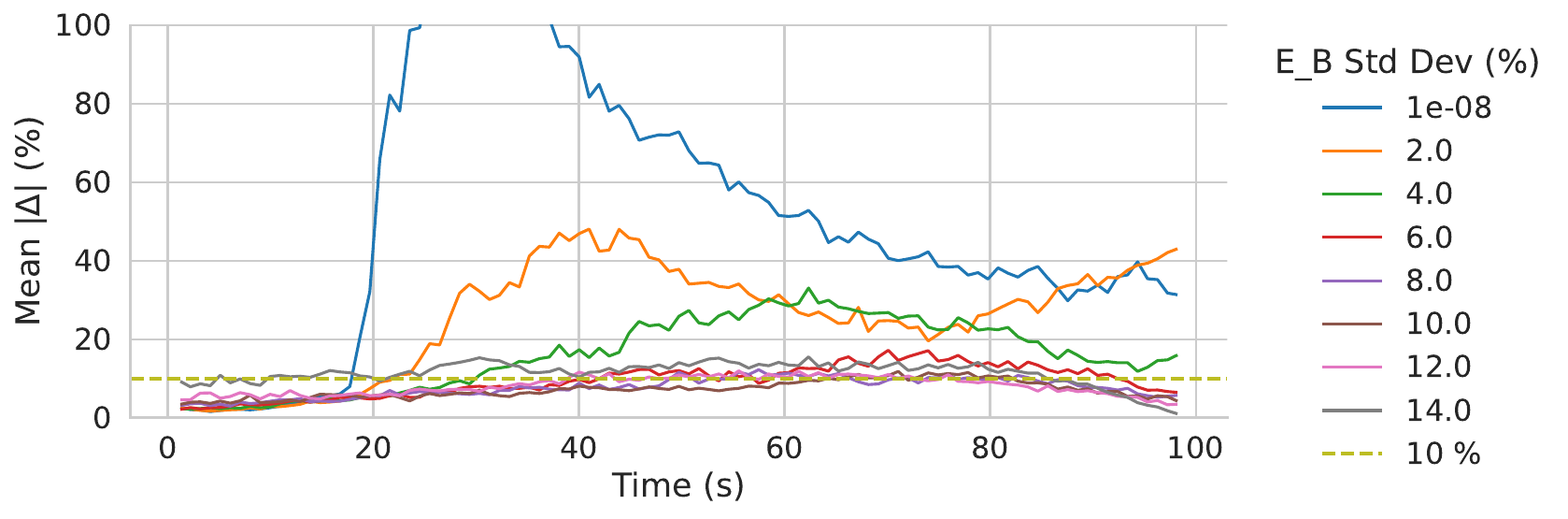}
    \Description{Line plot showing estimation error percentage versus elapsed time. Multiple colored lines represent different standard deviations of the energy barrier from 1 to 10 percent. Low variability yields minimal error at short times but limits range, while higher variability extends range with slightly increased error.}
    \caption{Impact of varying MTJ $E_b$ standard deviations.}
    \label{fig:eb23_eb_stddev}
\end{figure}

\begin{figure}[t]
    \centering
    \includegraphics[width=0.9\linewidth]{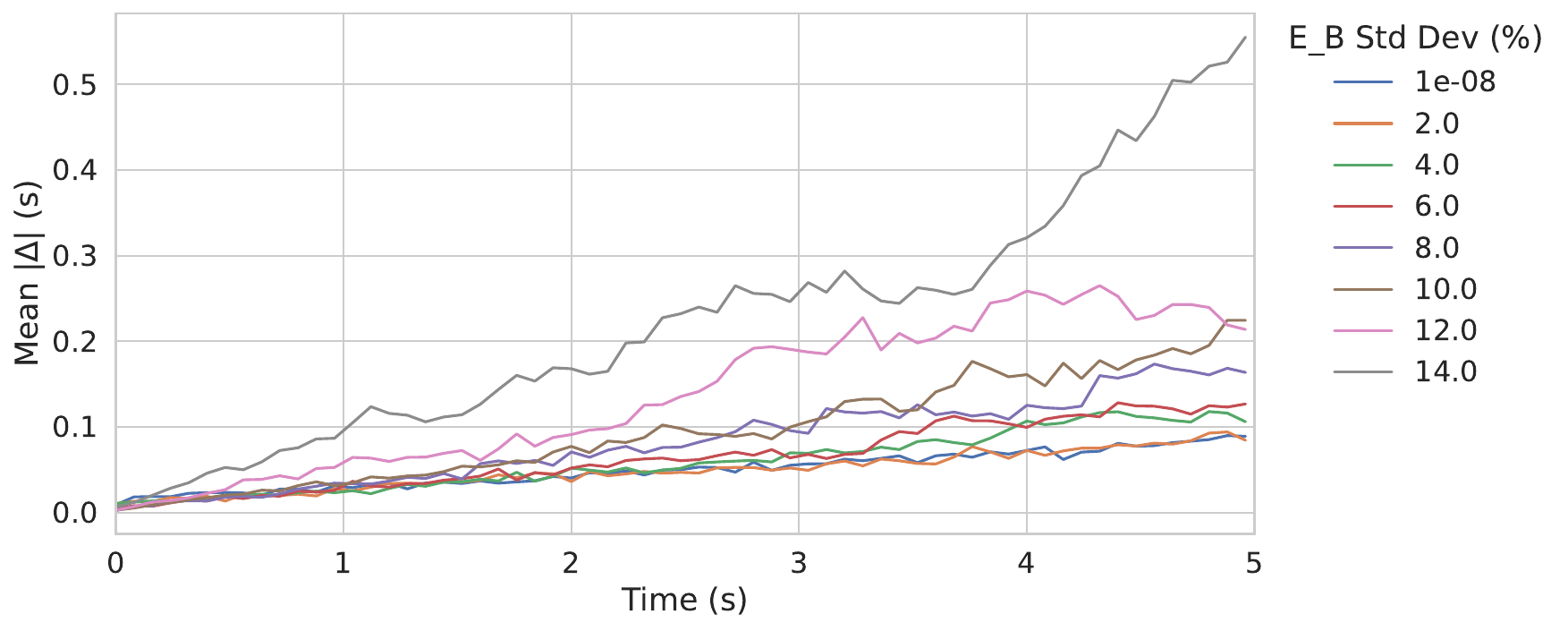}
    \Description{Line plot showing absolute time error in seconds versus elapsed time, focusing on early interval from 0 to 5 seconds. Multiple colored lines represent different energy barrier standard deviations. Shows how variation affects accuracy in initial measurement period.}
    \caption{Absolute error with varying MTJ $E_b$ standard deviations, in the $[0, 5]$ second interval.}
    \label{fig:eb23_eb_stddev_early_abs}
\end{figure}

\noindpar{Impact of $E_b$ Variability.} We study the impact of process-induced variation in the energy barrier $E_b$ by varying its standard deviation while keeping the mean constant (\autoref{fig:eb23_eb_stddev}). Low variability yields minimal error at short durations but limits the timekeeping range, as most MTJs switch at similar times. In contrast, higher variability spreads switching events over time, slightly increasing error but sustaining sub-10\% error for longer durations. We find that a 4–8\% standard deviation—consistent with typical MTJ fabrication—provides an effective operating point, balancing accuracy and range. This highlights a key insight: \sysname not only tolerates process variation, but benefits from it under realistic conditions.

\begin{figure}[t]
    \centering
    \includegraphics[width=\linewidth]{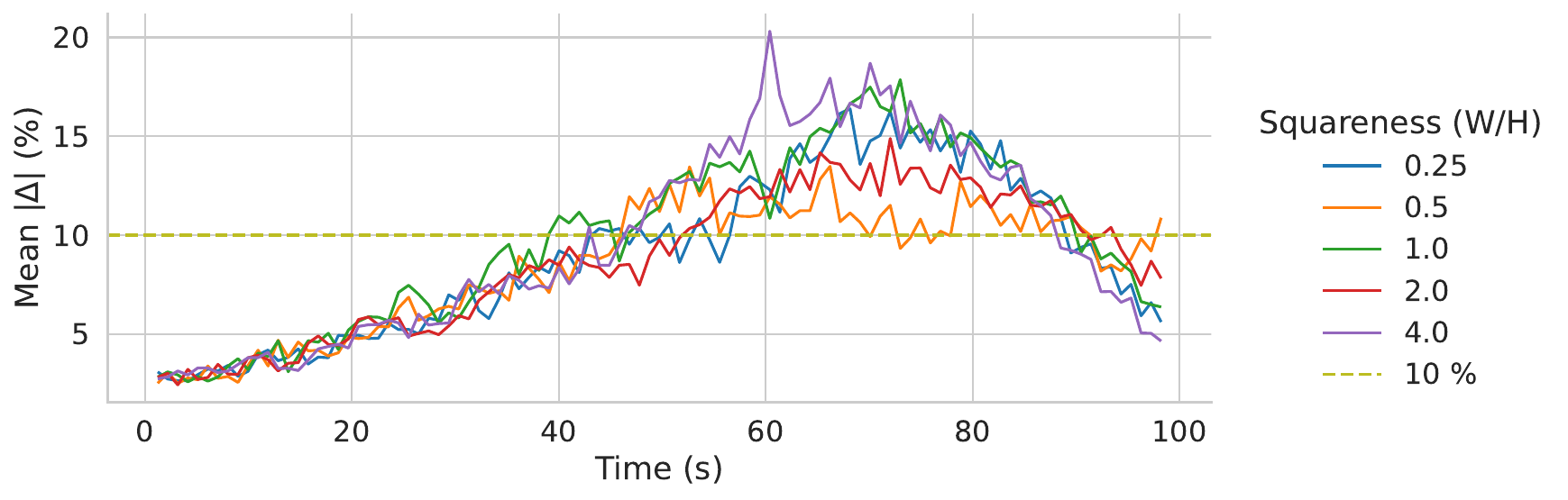}
    \Description{Line plot showing estimation error percentage versus elapsed time. Multiple colored lines represent different array aspect ratios defined by column to row ratio from 0.5 to 4. Squareness has minimal impact on error or timekeeping range.}
    \caption{
    Timekeeper error vs MTJ array squareness
    }
    \label{fig:eb23_squareness}
\end{figure}
\noindpar{Impact of Array Squareness.} We define the array squareness as the ratio of the number of columns to the number of rows. In \autoref{fig:eb23_squareness}, we plot the error as a function of squareness. We find that squareness does not impact the error or the timekeeping range. Thus, when reset voltage is a constraint, the array width may be reduced with very minimal impact.

\begin{figure*}
    \centering
    \begin{subfigure}{0.58\linewidth}
        \centering
        \includegraphics[width=\linewidth]{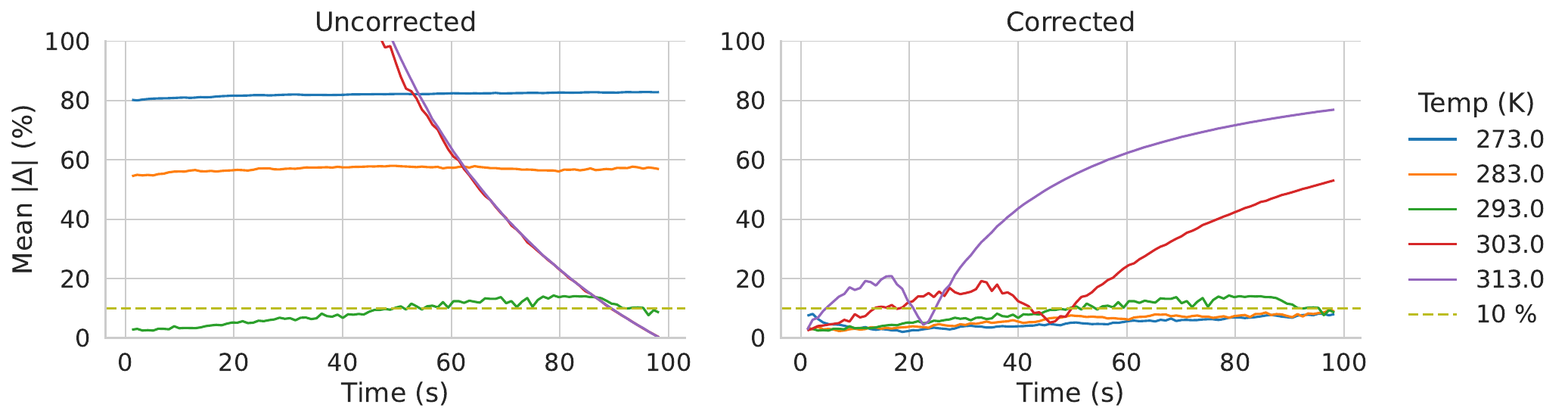}
        \Description{Line plot showing estimation error percentage versus elapsed time. Multiple colored lines represent different runtime temperatures from 273 to 313 Kelvin. Without correction, error increases substantially at higher temperatures, especially at longer times.}
        \caption{Only run temp variation.}
        \label{fig:eb23_temp_acc}
    \end{subfigure}
    ~
    \begin{subfigure}{0.42\linewidth}
        \centering
        \includegraphics[width=\linewidth]{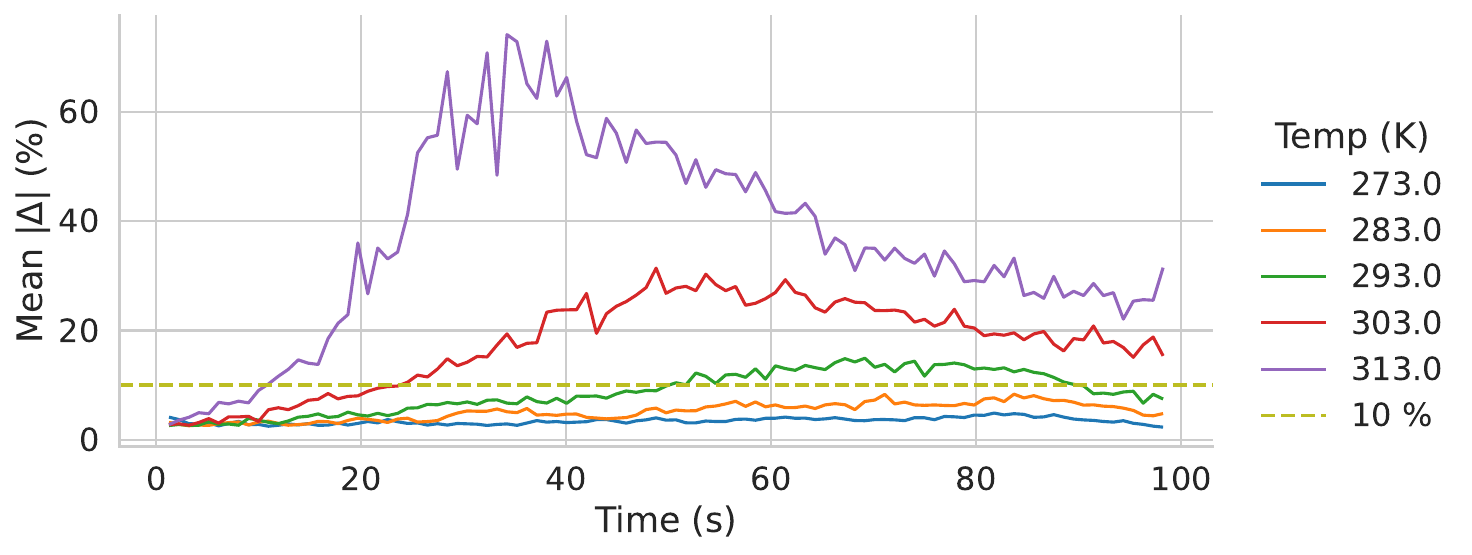}
        \Description{Line plot showing estimation error percentage versus elapsed time. Multiple colored lines represent different temperatures with calibration performed at each temperature. Shows improved consistency across temperature range compared to runtime-only correction.}
        \caption{Run and calibration temp variation.}
        \label{fig:eb23_temp_cal_acc}
    \end{subfigure}
    \caption{Timekeeper error with varying temperatures, at runtime and (optionally) calibration time.}
\end{figure*}

\noindpar{Impact of Temperature.} We evaluate the impact of runtime temperature on timekeeping accuracy in \autoref{fig:eb23_temp_acc}. Without correction, temperature variation degrades accuracy, especially at higher temperatures. Adding runtime temperature correction (\autoref{sec:temp-correction}) keeps error within 10\% in a $\pm\qty{20}{\K}$ range for a 40\,s interval, when physically possible (dwell times decrease at higher temperatures, limiting range). We further compare against calibration at different temperatures in \autoref{fig:eb23_temp_cal_acc}, where runtime and calibration temperatures are matched, achieving a similar 40\,s range with slightly lower error. These results show that lightweight runtime correction, enabled by widely available on-chip sensors~\cite{zhao_CMOSOnchipTemperature_2013} or calibration, is sufficient to maintain accuracy under realistic temperature variation.

\subsection{Multiple arrays}

A single array supports up to \qty{40}{\s} timekeeping within 10\% error. To extend this range, we compose four MTJ arrays and evaluate their combined behavior. 

\begin{figure*}
    \centering
    \begin{subfigure}{0.47\linewidth}
        \centering
        \includegraphics[width=\linewidth]{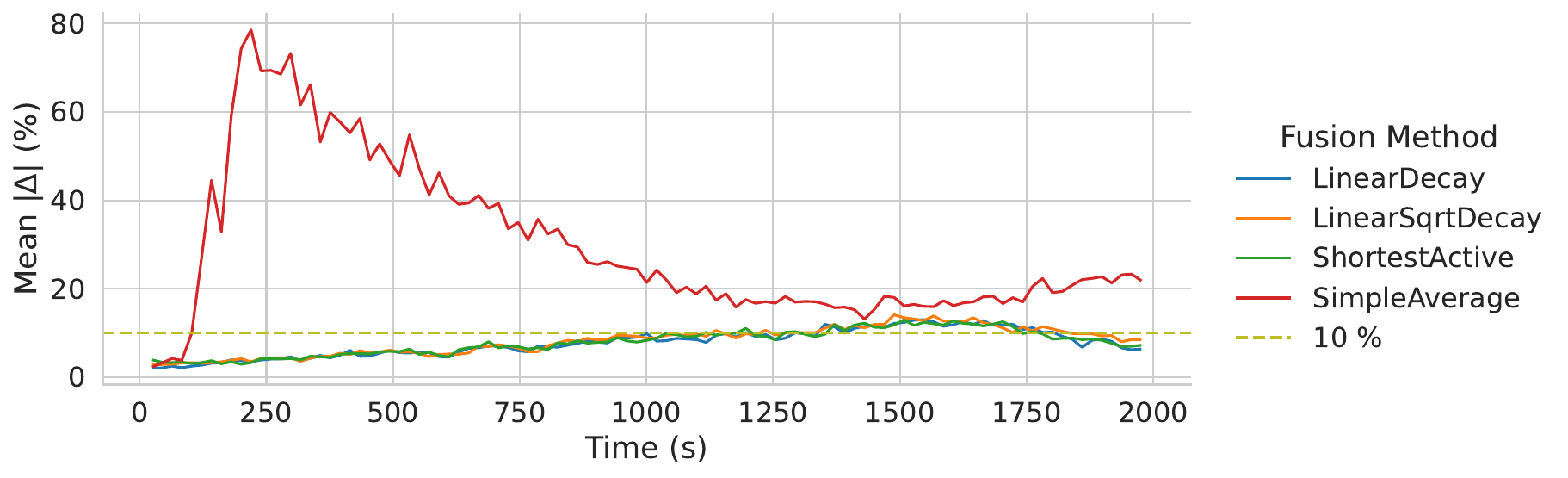}
        \Description{Line plot showing estimation error percentage versus elapsed time for four MTJ arrays. Multiple colored lines compare naive averaging against three weighted fusion methods. Naive averaging fails quickly, while weighted methods maintain accuracy across the full range.}
        \caption{Fusion Methods.}
        \label{fig:multi_fusion_method}
    \end{subfigure}
    ~
    \begin{subfigure}{0.53\linewidth}
        \centering
        \includegraphics[width=\linewidth]{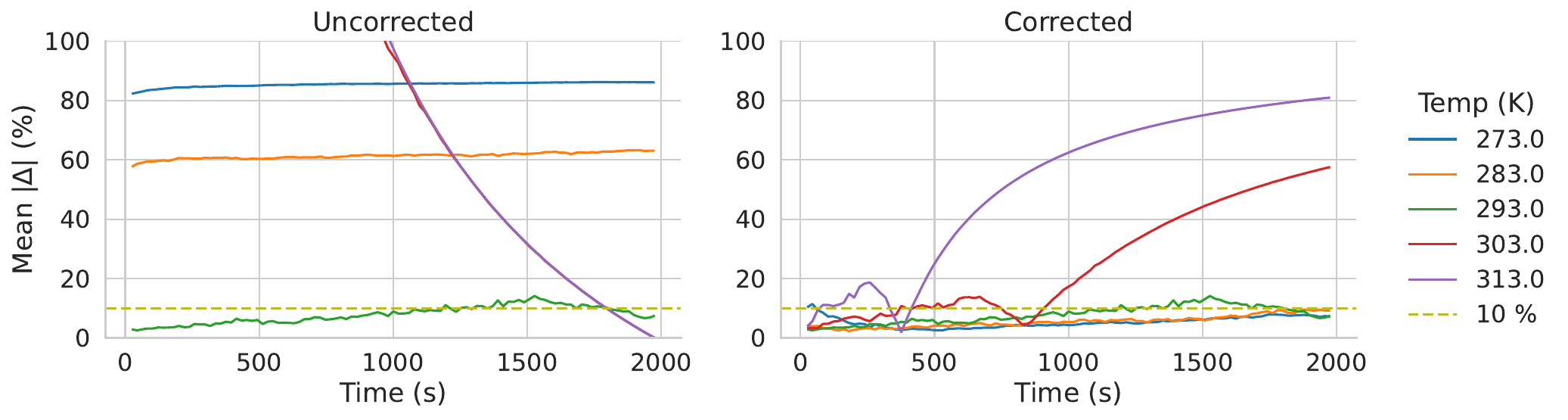}
        \Description{Line plot showing estimation error percentage versus elapsed time for four array system. Multiple colored lines represent different runtime temperatures from 273 to 313 Kelvin. Runtime temperature correction maintains error below 10 percent (where possible) across the operating range.}
        \caption{Run Temperature Variation.}
        \label{fig:multi_temp_acc}
    \end{subfigure}
    \caption{Multiple array timekeeper error with varying fusion methods and temperature variations.}
    \label{fig:multi_results}
\end{figure*}

MTJ dwell times and corresponding energy barriers follow stochastic distributions, which differ across arrays. Na\"ive averaging fails to capture these dynamics, as it assumes a stationary and identical distribution across arrays. As shown in \autoref{fig:multi_fusion_method}, this results in high error, rendering the approach unreliable.
In contrast, fusion methods that assign adaptive weights to observations—particularly those that prioritize more recent data—better capture the evolving system dynamics and significantly improve stability. All three methods achieve a 938\,s range within 10\% error; we use Linear-Decay Sqrt as the default, as it achieves the lowest error on average. 

We evaluate robustness under a $\pm20\,K$ temperature variation in \autoref{fig:multi_temp_acc}. Runtime temperature correction using 4 arrays also achieves a \qty{938}{\s} range under 10\% error when physically possible.


\subsection{Energy, Time, and Space Overheads}
\label{sec:results-energy}



\noindpar{Energy:} Using a Cadence MTJ simulator~\cite{garcia-redondo_CompactModelScalable_2021, garcia-redondo_FokkerPlanckSolverModel_2021}, we determine that a \qty{50}{\ns}, \qty{35}{\uA} pulse reliably resets a single MTJ. Thus, in the worst case (AP state, $R = \qty{118.26}{\kilo\ohm}$), the energy needed to reset a single MTJ is $E_{reset} = I^2 \cdot R\cdot t  = \qty{7.24}{\pico\joule}$.
Thus, with four $64 \times 256$ arrays (\num{65536} devices) the total required energy is $\widehat{E_{reset}} = \qty{474.7}{\nano\joule}$.

Then, to measure the MTJs, there are three energy costs: array/shunt resistor, ADC, and software. The energy dissipated in the array and shunt resistor in the worst case (P state, $R_{array} = \qty{13.2}{\kilo\ohm}$) can be easily calculated. Assuming an ADC sample time of \qty{1}{\us}~\cite{texasinstruments_MSP430FR596xMSP430FR594xMixedSignal_2018}, the total energy comes to $\widehat{E_{meas}} = \qty{4.28}{\nano\joule}$. The ADC energy is \qty{87}{\nano\joule} per channel (experimentally measured on a MSP430FR5994~\cite{texasinstruments_MSP430FR596xMSP430FR594xMixedSignal_2018}) for a total of $E_{ADC} = \qty{348}{\nano\joule}$ across the four arrays. 

To estimate software energy, we utilized \texttt{llvm-mca}~\cite{llvmproject_LlvmmcaLLVMMachine_2026}. Since \texttt{llvm-mca} does not support the MSP430 ISA, we target an ARM Cortex-A15~\cite{arm_CortexA15MPCoreProcessor_2013} with clang++ (commit \texttt{72df1fc6});
the Cortex-A15 is higher-power than typical batteryless MCUs, giving us a conservative upper bound on software energy.
We determine that the software time estimation code takes 401 cycles to complete. Using the energy estimations in~\cite{vasilakis_InstructionLevelEnergy_2015}, total software energy is estimated as $E_{sw} = \qty{207.74}{\nano\joule}$. Thus, total \sysname system energy use is
\begin{equation}
    E_{total} = \widehat{E_{reset}} + \widehat{E_{meas}} +E_{ADC} +  E_{sw} = \qty{1034.72}{\nano\joule}
\end{equation}


\noindpar{Time:} At the MSP430's minimum \qty{6}{\mega\Hz}, 401 cycles take \qty{66.83}{\us}, a conservative estimate; adding four \qty{1}{\us} ADC samples~\cite{texasinstruments_MSP430FR596xMSP430FR594xMixedSignal_2018} gives \qty{70.83}{\us} total startup overhead. Shutdown time overhead is the reset pulse length: \qty{50}{\us}.

\noindpar{Memory:} Each array has a calibration table with 128 entries of two 4-byte floats each, totaling \qty{1024}{\byte} of space. Thus, four tables need \qty{4}{\kibi\byte}, which easily fits in the $\approx\qty{512}{\kibi\byte}$ of non-volatile memory found in common MCUs~\cite{espressifsystems_ESP32WROOM32Datasheet_2025, texasinstruments_MSP430FR596xMSP430FR594xMixedSignal_2018}. If we only store time \textit{values}, and compute the LinSpace (\autoref{alg:calibration}) every startup, this can be reduced to \qty{2}{\kibi\byte} for the tables, and an additional \qty{12}{\byte} for LinSpace parameters, totaling \qty{2060}{\byte}. However, this incurs additional compute overheads.


\noindpar{Size:} An MTJ is \qty{200}{\nm} in diameter, including spacing to prevent magnetic coupling~\cite{zhang2011low}. Thus, a $64 \times 256$ array requires \qty{655.36}{\um^2}. Within \qty{0.1}{\mm^2}, we could fit 100 arrays!


\subsection{Comparison of \sysname with  the SOTA}

\begin{table}[h!tbp]
	\footnotesize
	\centering
 	\caption{Comparison of \sysname with SOTA timekeepers.
    }
	\begin{tabular}{ m{15mm} m{3.8mm} m{3.8mm} m{3.8mm} m{7mm} m{7mm} m{8.5mm} m{8.5mm}}
		\toprule
		~ & \textbf{C1} & \textbf{C2}& \textbf{C3} & \textbf{CHRT}  & \textbf{HARC} & \textbf{AM1805} & \cellcolor{babyblueeyes!10} \textbf{\sysname}\\

		\midrule
		
		 \rowcolor{black!3} 
		\textbf{Range (s)} & 0.91 & 9.13 & 91.3 & 102 & 91.33 & 110 &\cellcolor{babyblueeyes!10} 273.6 \\
		
		
        

        \textbf{$I_{static}$ (nA) } & 30 & 30 & 30 & 958 & 0 & 55 & \cellcolor{babyblueeyes!10} 0 \\

        \rowcolor{black!3} 
        \textbf{$E_{timer}$ (uJ)} & 0.098 & 0.98 & 9.85 & 10.94 &  10.94  & 0  &  \cellcolor{babyblueeyes!10}0.356 \\ 
        
        \textbf{$E_{ADC}$ (nJ)} & 87 & 87  & 87  & 261 & 261  & 0 &  \cellcolor{babyblueeyes!10}261 \\   
        
        
        \rowcolor{black!3} 
        \textbf{$t_{startup}$ (ms)} & 5.4 & 54.6 & 546.7 & 607.24 & 607.24  & 900 &  \cellcolor{babyblueeyes!10} 0.071 \\   
        
        \textbf{Resolution (s)} & 0.01 & 0.1 & 1 & 0.01-1  & 0.01-1  & 0.01 &  \cellcolor{babyblueeyes!10} 0.005-1 \\    

        \rowcolor{black!3}
        \textbf{NVM Use (B)} & 8192 & 8192 & 8192 & 24576 & 11001 & N/A & \cellcolor{babyblueeyes!10}24576\\
        
        \textbf{Size (mm\textsuperscript{2})} & 2.5 & 2.5 & 2.5 & 7.5 & 7.5 & 9 & \cellcolor{babyblueeyes!10} 0.002\\

        

		\bottomrule
	\end{tabular}

	\label{table:SOA_comp}
\end{table}

We summarize a comparison of \sysname against prior timekeepers in \autoref{table:SOA_comp}, including an Ambiq AM1805 RTC~\cite{ambiq_am1805_datasheet}. The $E_{timer}$ row reflects the cost to charge or reset the timing element only; ADC and software costs are itemized separately and apply equally to all approaches (\autoref{sec:results-energy}). \sysname significantly extends the maximum timekeeping range to \qty{273.6}{\s}, achieving $3.00\times$ and $2.68\times$ longer durations than HARC and CHRT, respectively. At the same time, it achieves time resolution as low as \qty{5}{\ms}, while maintaining a worst-case resolution of \qty{1}{\s}---equivalent to prior approaches. When the range is limited to \qty{100}{\s}, the worst-case resolution can be improved to \qty{0.19}{\s}.
Unlike prior work, \sysname needs no dedicated capacitors or active circuitry for off-time measurement, eliminating startup latency and active current. The only required operation is one ADC measurement per array after each power interruption,  similar to prior approaches, which require one measurement per capacitor.

We further evaluate time-estimation accuracy using real energy-harvesting traces (\autoref{table:CHRTvsFLINT}). CHRT is unable to reliably track time under long and highly variable off-time conditions, as its charging-time-based estimation cannot cover extended interruptions, whereas \sysname maintains accurate estimation. Both methods achieve similar accuracy under stable conditions, but \sysname significantly extends the operating range and enables more applications with its lower energy use, while preserving estimation quality.

\begin{table*}[h!tbp]
    \footnotesize
    \centering
    \caption{Estimation error (\%) for \sysname vs.\ CHRT. CHRT has startup time before execution. DNF = capacitor could not charge.}
    \begin{tabular}{lcccccccccccc}
\toprule
\textbf{} & \multicolumn{3}{c}{\cellcolor{black!3}\textbf{\textsf{Trace A}}} & \multicolumn{3}{c}{\textbf{\textsf{Trace B}}} & \multicolumn{3}{c}{\cellcolor{black!3}\textbf{\textsf{Trace C}}} & \multicolumn{3}{c}{\textbf{\textsf{Trace D}}} \\ 
\midrule 
\textbf{Sys. Cap. Size} & \cellcolor{black!3}\textbf{\textsf{10 uF}} & \cellcolor{black!3}\textbf{\textsf{100 uF}} & \cellcolor{black!3}\textbf{\textsf{1 mF}} & \textbf{\textsf{10 uF}} & \textbf{\textsf{100 uF}} & \textbf{\textsf{1 mF}} & \cellcolor{black!3}\textbf{\textsf{10 uF}} & \cellcolor{black!3}\textbf{\textsf{100 uF}} & \cellcolor{black!3}\textbf{\textsf{1 mF}} & \textbf{\textsf{10 uF}} & \textbf{\textsf{100 uF}} & \textbf{\textsf{1 mF}} \\ 
\textbf{Avg. Off Time (s)} & \cellcolor{black!3}1.55 & \cellcolor{black!3}15.22 & \cellcolor{black!3}190.54 & 2.86 & 12.62 & 168.22 & \cellcolor{black!3} 0.47 & \cellcolor{black!3}4.35 & \cellcolor{black!3} 45.24  & 5.03 & 51.09 & \textcolor{red}{DNF} \\
\midrule
\textbf{CHRT} & \cellcolor{black!3} 2.25\% & \cellcolor{black!3} 
{\color{red}Failed} & \cellcolor{black!3} \textcolor{red}{Failed} & 0.83\% & 3.14\% & \textcolor{red}{Failed} & \cellcolor{black!3}0.71\% & \cellcolor{black!3} 2.94\% & \cellcolor{black!3} \textcolor{red}{Failed} & 3.32\% & \textcolor{red}{Failed} & --- \\ 
\textbf{FLINT (\%)} & \cellcolor{black!3}1.69\% & \cellcolor{black!3}2.52\% & \cellcolor{black!3}3.09\% & 1.60\% & 1.34\% & 1.32\% & \cellcolor{black!3}1.64\% & \cellcolor{black!3}1.07\% & \cellcolor{black!3} 1.49\% & 2.05\% & 2.40\% & --- \\ 
\bottomrule
\end{tabular}
\label{table:CHRTvsFLINT}
\end{table*}



\subsection{Long-Term Reliability Under Aging}
\label{sec:results-aging}

\begin{figure*}
    \centering
    \begin{subfigure}[t]{0.33\linewidth}
        \centering
        \includegraphics[width=0.99\linewidth]{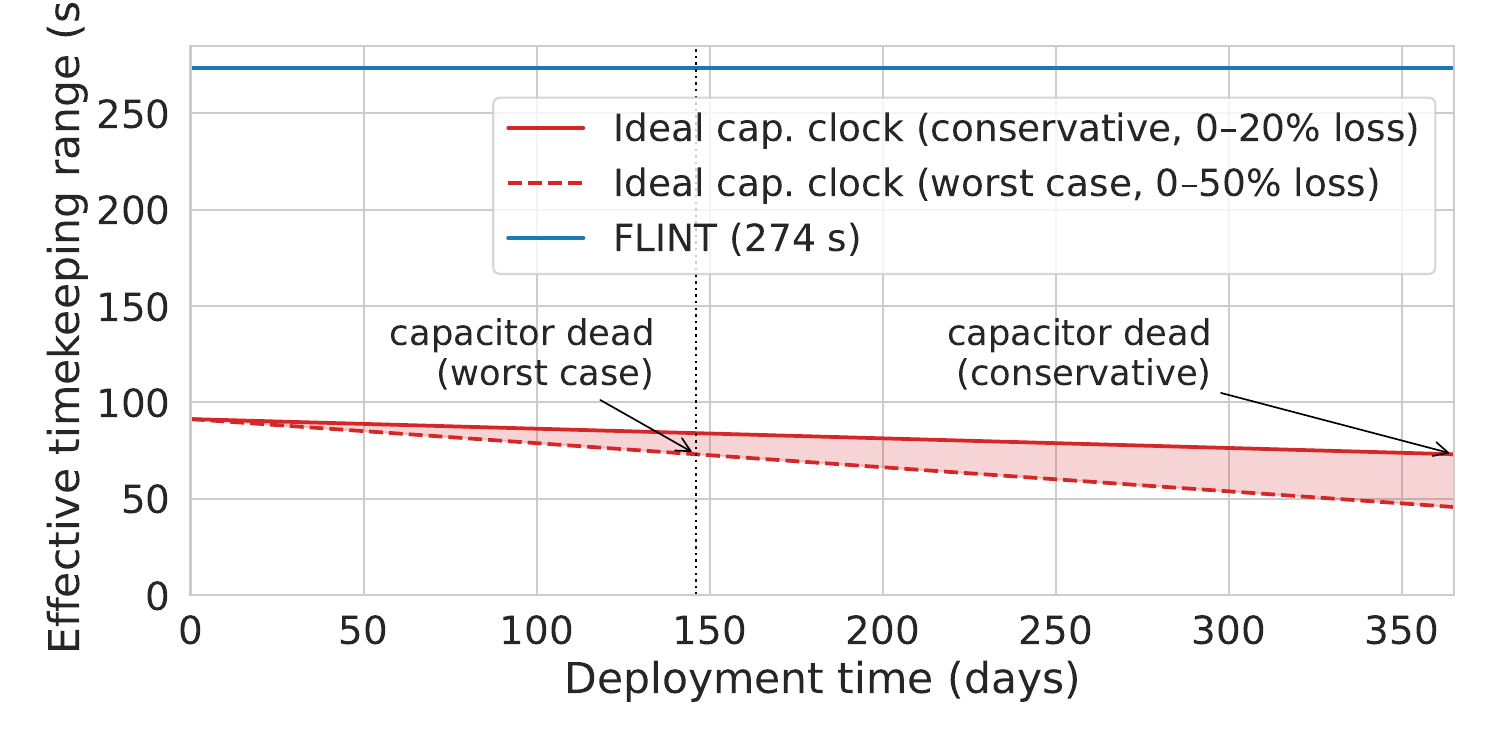}
        \Description{Line plot showing maximum timekeeping range in seconds versus days over one year. FLINT system maintains constant 273.6 second range. Ideal capacitor starts at 91.3 seconds and degrades to 73 seconds (conservative) or 45.7 seconds (worst-case).}
        \caption{Effective range over one year.}
        \label{fig:aging_range}
    \end{subfigure}
    ~
    \begin{subfigure}[t]{0.33\linewidth}
        \centering
        \includegraphics[width=0.99\linewidth]{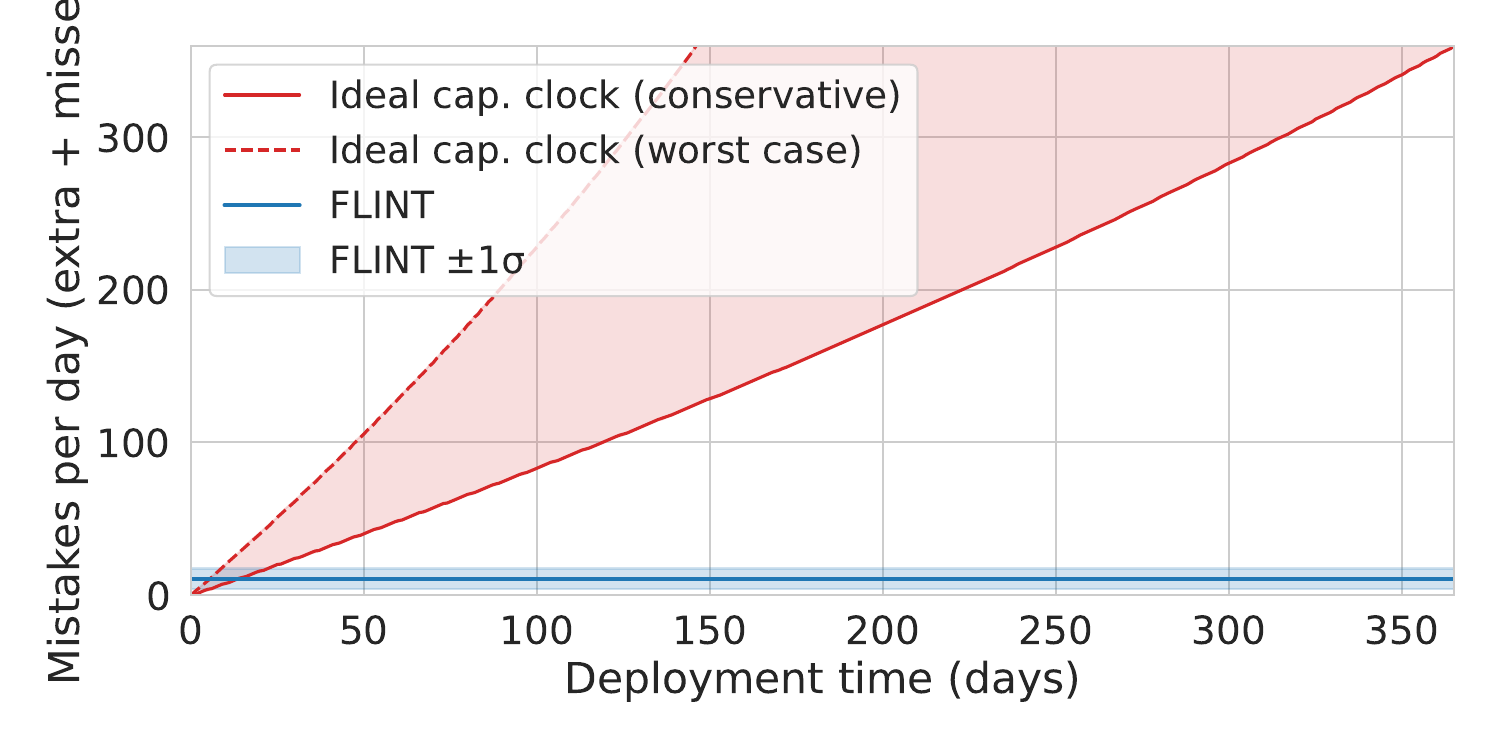}
        \Description{Bar or line plot showing number of scheduling mistakes per day for a 60-second periodic task over one year. FLINT shows minimal mistakes around 2000 per year. Ideal capacitor shows zero mistakes initially then increases to tens of thousands per year as aging bias accumulates.}
        \caption{Daily scheduling mistakes (\qty{60}{\s} task).}
        \label{fig:aging_expirations}
    \end{subfigure}
    ~
    \begin{subfigure}[t]{0.33\linewidth}
        \centering
        \includegraphics[width=0.99\linewidth]{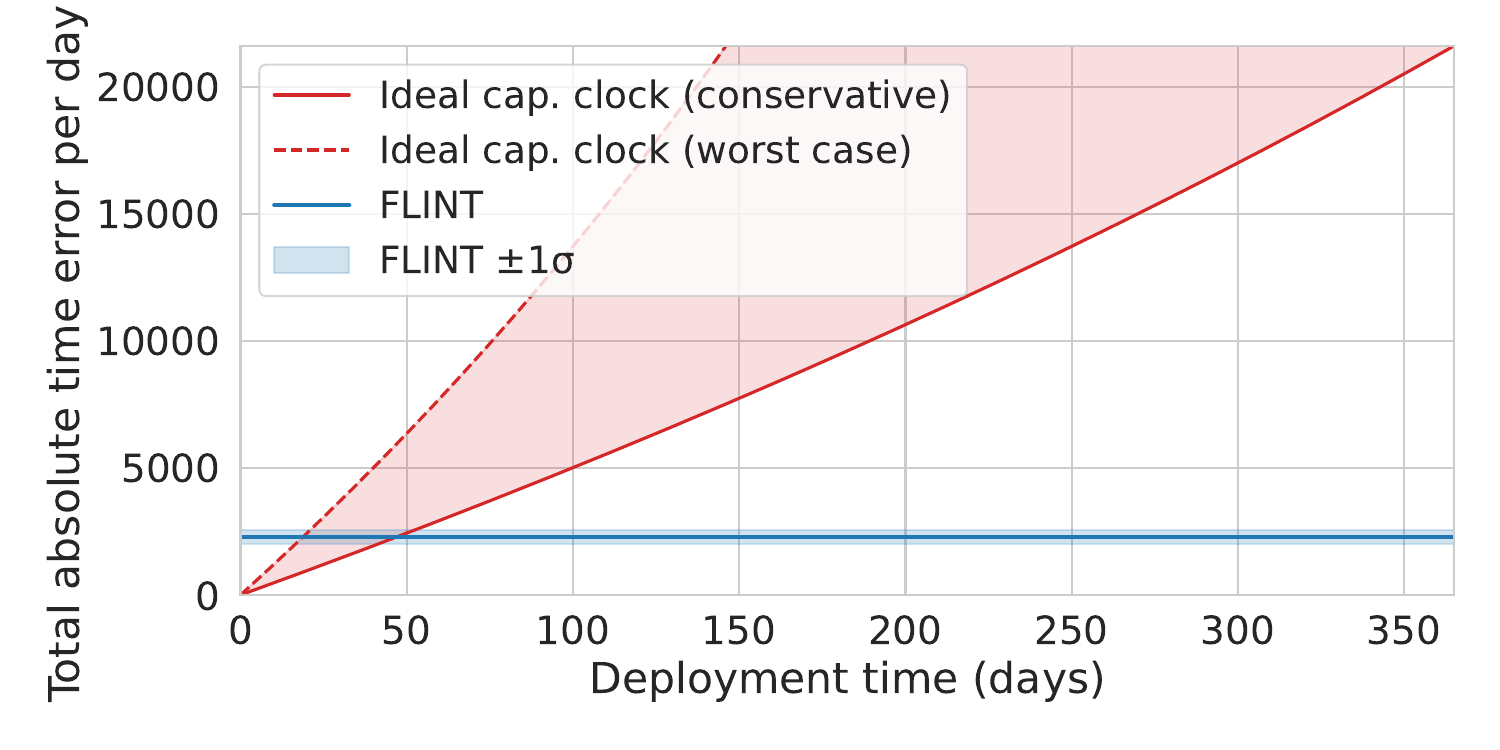}
        \Description{Line plot showing accumulated absolute time error in seconds versus days over one year. FLINT maintains nearly flat error accumulation. Ideal capacitor error grows rapidly in first two months then continues accumulating to thousands of seconds per day by year end.}
        \caption{Absolute time error accumulated per day.}
        \label{fig:aging_time_error}
    \end{subfigure}
    \caption{\sysname vs.\ ideal capacitor clock over one year. Bands: conservative (0--20\%/yr) to worst-case (0--50\%/yr) degradation~\cite{choi2022capos}.}
    \label{fig:aging}
\end{figure*}

The comparison above assumes a fresh capacitor; \autoref{fig:aging} shows how aging erodes that baseline over one year.

\noindpar{Range.} \sysname holds \qty{273.6}{\s} for the device's life, while the ideal capacitor clock falls from \qty{91.3}{\s} to \qty{73.0}{\s} (conservative) or \qty{45.7}{\s} (worst case) within a year, crossing a 20\% loss ``capacitor dead'' threshold at day 365 and day 146 respectively. \sysname's $3.0\times$ range advantage at deployment grows monotonically thereafter.

\noindpar{Accuracy and scheduling.} The accuracy cost is sharper than the range cost. Using the re-sample mistake metric defined in \autoref{sec:eval-aging} (fixed \qty{60}{\s} scheduled task), the ideal capacitor clock's aging bias has a fixed sign---it always reads high and fires early---so the error does not average out but accumulates across the thousands of cycles in a day (\autoref{fig:aging_expirations}). Averaged over a year, the ideal capacitor clock incurs \num{60889} redundant samples (conservative) to \num{203659} (worst case), versus roughly \num{1992} redundant and \num{1889} missed for \sysname, whose errors are zero-mean and do not accumulate. This is a $16\times$ to $52\times$ reduction. The same one-directional bias inflates the total mistimed duration: summed over a day, the ideal capacitor clock's absolute accumulated time error grows with age and overtakes \sysname's flat error within roughly the first two months of deployment (\autoref{fig:aging_time_error}).

\subsection{Application Case Studies}
\label{sec:results-casestudies}
\label{sec:eval-casestudies}

\begin{table*}[t]
\footnotesize
\centering
\caption{Case study configurations. Note \textnormal{$t_{\textit{off}} = b + s \cdot \text{Bin}(n, 0.5)$}, and \textnormal{$\overline{t_{\textit{off}}}$\,=\,mean off time.}}
\label{tab:cs_configs}
\begin{tabular}{l l c c c c l c}
\toprule
\textbf{ID} & \textbf{Application} & $b$ & $s$ & $n$ & $\overline{t_{\textit{off}}}$ & \textbf{Mistake definition} & \textbf{$E_b$ (kT)} \\
\midrule
\rowcolor{black!3}
A1 & Position logging~\cite{juang2002zebranet,szewczyk2004habitat,markham2010magnetoinductive} & \qty{1.0}{\s} & \qty{0.2}{\s} & 30 & \qty{4}{\s} & Early fire: true elapsed $<\qty{9}{\s}$; stale data: true age $>\qty{66}{\s}$ & 23--26 \\
A2 & Structural health monitoring~\cite{afanasov2020battery} & \qty{10}{\s} & \qty{1}{\s} & 50 & \qty{35}{\s} & Sample outside $[\qty{810}{\s},\qty{990}{\s}]$ ($\pm10\%$ of $T^*{=}\qty{900}{\s}$) & 23--26 \\
\rowcolor{black!3}
A3 & Cold-chain monitoring~\cite{mercier2017coldchain} & \qty{10}{\s} & \qty{1}{\s} & 30 & \qty{25}{\s} & False condemn ($\hat{E}{>}1.1B$, safe) or missed condemn ($\hat{E}{\le}1.1B$, spoiled) & 23--26 \\
\midrule
D4 & Overnight solar gap & \multicolumn{4}{c}{target: $\approx\qty{12}{\hour}$} & Geomean \% error (no cap baseline) & 28--31 \\
\rowcolor{black!3}
D5 & Far-future wakeup & \multicolumn{4}{c}{target: $\approx\qty{7}{\day}$} & Geomean \% error (no cap baseline) & 31--34 \\
\bottomrule
\end{tabular}
\end{table*}

We evaluate \sysname on five representative intermittent computing applications, demonstrating its versatility across timing regimes that capacitor clocks either handle poorly or cannot reach at all.

\begin{itemize}
    \item \textit{A1 Position logging}~\cite{juang2002zebranet,szewczyk2004habitat,markham2010magnetoinductive}: a batteryless GPS tag fires periodic fixes to log wildlife trajectories; timing accuracy is critical to avoid wasted \qty{9}{\milli\joule} GPS acquisitions and stale position data.
    \item \textit{A2 Structural health monitoring}~\cite{afanasov2020battery}: a batteryless sensor on a bridge samples vibration at regular intervals; capacitor aging introduces cumulative drift that biases the collected data.
    \item \textit{A3 Cold-chain monitoring}~\cite{mercier2017coldchain}: a batteryless tag accumulates unsafe-temperature exposure; directional aging bias of capacitors leads to wrong safety verdicts on shipments.
    \item \textit{D4 Overnight solar gap}: a solar-powered node must estimate time elapsed at dawn to decide whether a buffered reading is still fresh; the \qty{12}{\hour} gap exceeds any practical capacitor.
    \item \textit{D5 Far-future wakeup}: a wildlife tag must schedule a weekly radio transmission; a \qty{7}{\day}\ off-time is impossible to measure with any capacitor.
\end{itemize}

The first three (A1--A3) prioritize \emph{measurement accuracy}: off-times fall within the capacitor clock's range, enabling direct comparison against an aging capacitor baseline (C3 from \autoref{sec:eval-aging}). Off-times are drawn from a shifted-binomial model $t = b + s \cdot \mathrm{Bin}(n,\, 0.5)$ with means below the C3 baseline's $\sim\qty{91}{\s}$ ceiling, isolating aging bias from range effects. The next two (D4--D5) prioritize \emph{measurement range}: the required off-times exceed the practical capacitor ceiling, so we report \sysname error at the endpoint alongside the capacitor energy and area that \emph{would} be required.
\sysname uses the default multi-array configuration from \autoref{sec:eval-arr-sims}; $E_b$ overrides are in \autoref{tab:cs_configs}. Each application defines a $\pm\qty{10}{\percent}$ tolerance on its target interval $T^*$; mistake rates are the arithmetic mean across 20 device realizations.

\noindpar{Accuracy analysis (\autoref{tab:cs_accuracy}).}
Capacitor aging introduces an ever-increasing timing bias. The capacitor makes zero mistakes until this bias exceeds \qty{10}{\percent}---day~183 ($\alpha{=}0.2$) or day~74 ($\alpha{=}0.5$)---after which decisions become systematically wrong. \sysname's errors carry no consistent bias and do not accumulate with deployment age.

\begin{table}[t]
\footnotesize
\centering
\caption{Accuracy case study mistakes per year.}
\label{tab:cs_accuracy}
\begin{tabular}{l c c}
\toprule
\textbf{Application} & \cellcolor{babyblueeyes!10}\textbf{\sysname/yr}
    & \textbf{Opt.\ cap/yr (con.--wst.)} \\
\midrule
\rowcolor{black!3}
Position logging   & \cellcolor{babyblueeyes!10}\textbf{0} & 1.86M -- 3.71M \\
Struct.\ health    & \cellcolor{babyblueeyes!10}\textbf{0} & 20.7K -- 41.2K \\
\rowcolor{black!3}
Cold-chain         & \cellcolor{babyblueeyes!10}\num{18} (\textbf{0 tuned}) & \numrange{117}{266} \\
\bottomrule
\end{tabular}
\end{table}

\textit{A1}:
\sysname makes zero early-fire or stale-data mistakes across all 20 trials.
The aging capacitor fires every GPS fix more than \qty{1}{\s} early from day~\numrange{74}{183} onward,
accumulating \numrange{1.86}{3.71}~million premature \qty{9}{\milli\joule} fixes per year
(\qty{35}{\kilo\joule\per\syear} wasted~\cite{markham2010magnetoinductive}).

\textit{A2}:
\sysname has zero mis-timed samples across all 20 device trials.
The aging capacitor mis-schedules \numrange{20700}{41200} samples per year.

\textit{A3}:
At the default $1.1B$ condemn threshold, \sysname
makes zero false condemnations; 1/20 device realizations miss a spoiled shipment
(\num{18}/yr). Lowering the threshold to $B$ eliminates all missed condemns with zero
false condemns---an application-level tuning with no hardware change.
The aging capacitor falsely condemns safe shipments from
day~\numrange{100}{249} onward (\numrange{117}{266} wrong verdicts per year)---a one-sided, growing failure mode absent
in \sysname.

\begin{table}[t]
\footnotesize
\centering
\caption{Duration case study results.}
\label{tab:cs_duration}
\begin{tabular}{l c r r r}
\toprule
\textbf{Application} & \textbf{Range} & \cellcolor{babyblueeyes!10}\textbf{\sysname err.} & \textbf{Cap. energy} & \textbf{Cap. size}\\
\midrule
\rowcolor{black!3}
Overnight solar gap & \qty{12}{\hour} & \cellcolor{babyblueeyes!10}\qty{2.2}{\percent} & \qty{4.7}{\milli\joule} & \qty{1.0}{\milli\farad}\\
Far-future event & \qty{7}{\day} & \cellcolor{babyblueeyes!10}\qty{2.3}{\percent} & \qty{65.2}{\milli\joule} & \qty{14.6}{\milli\farad}\\
\bottomrule
\end{tabular}
\end{table}


\begin{figure}[t]
    \centering
    \includegraphics[width=\linewidth]{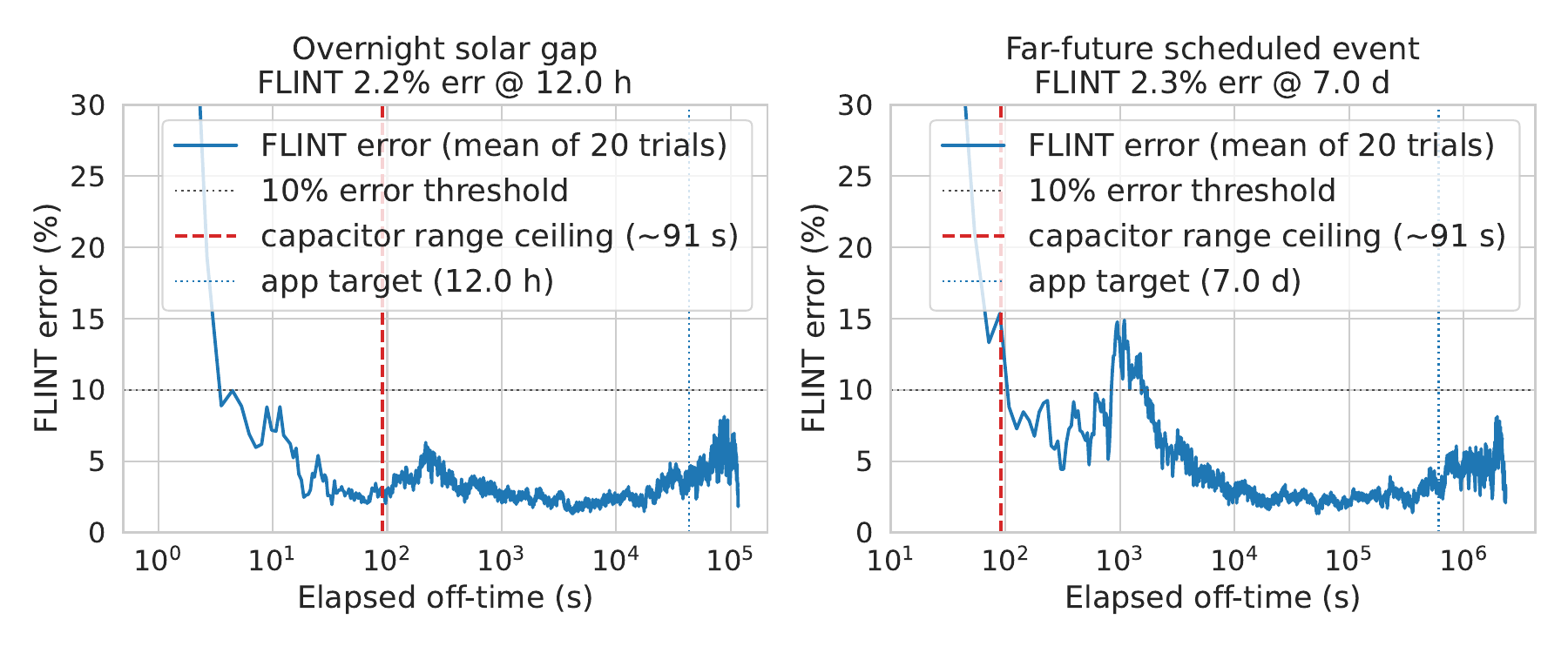}
    \Description{Line plot showing FLINT estimation error percentage versus elapsed time on logarithmic scale. Demonstrates under 10 percent error sustained across hours to days range, far beyond the capacitor clock range ceiling of 91 seconds.}
    \caption{\sysname error in duration case studies, reaching hours-to-days ranges at $\sim$\qty{2}{\percent} err---impossible for a capacitor.}
    \label{fig:cs_duration}
\end{figure}

\begin{table*}[htbp]
    \footnotesize
    \centering
    \caption{\small Comparison of timekeeping approaches across key design axes. 
    Colors:     \textcolor{green!55!black}{Fully satisfies} $|$ 
    \textcolor{orange!80!black}{Partially satisfies} $|$
    \textcolor{red!70!black}{Does not satisfy}.
    }
    \begin{tabular}{l c c c c c c c c c}
        \toprule
        \textbf{System} &
        \textbf{Category} &
        \textbf{\makecell{Time \\ Source}} &
        \textbf{\makecell{Aging \\ Sensitivity}} &
        \textbf{\makecell{Environmental \\Robustness}} &
        \textbf{Resolution} &
        \textbf{\makecell{Max \\ Range}} &
        \textbf{\makecell{{Energy vs.}\\{Range}}} &
        \textbf{\makecell{Startup \\ Overhead}} &
        \textbf{\makecell{Reconfig \\ Required?}} \\
        \midrule

        \textbf{AM1805~\cite{ambiq_am1805_datasheet}}
        & HW & RTC
        & \textcolor{orange!80!black}{Medium (BCA)}
        & \textcolor{orange!80!black}{Medium (BCA)} 
        & \textcolor{green!55!black}{High}
        & \textcolor{green!55!black}{High}
        & \textcolor{red!70!black}{Linear}
        & \textcolor{red!70!black}{Very High}
        & \textcolor{red!70!black}{Yes} \\

        \rowcolor{black!3}
        \textbf{TARDIS~\cite{hester2016persistent}}
        & SW & SRAM
        &  \textcolor{green!55!black}{None}
        & \textcolor{red!70!black}{Low (Temperature)}
        & \textcolor{red!70!black}{Low}
        & \textcolor{red!70!black}{Low}
        & \textcolor{orange!80!black}{Const (Low Range)}
        & \textcolor{green!55!black}{Low}
        & \textcolor{green!55!black}{No} \\

        \textbf{Osc.-based~\cite{alsubhi2020can}}
        & SW & Osc.
        &  \textcolor{green!55!black}{None}
        & \textcolor{red!70!black}{Low}
        & \textcolor{green!55!black}{High}
        & \textcolor{red!70!black}{Very Low}
        & \textcolor{orange!80!black}{Const (Low Range)}
        & \textcolor{green!55!black}{Low}
        & \textcolor{green!55!black}{No} \\

        \rowcolor{black!3}
        \textbf{CUSTARD~\cite{hester2016persistent}}
        & HW & Cap.
        & \textcolor{red!70!black}{Cap. Drift}
        & \textcolor{red!70!black}{Low}
        & \textcolor{green!55!black}{High}
        & \textcolor{red!70!black}{Low}
        &  \textcolor{red!70!black}{Linear}
        & \textcolor{red!70!black}{ Scales w/ Cap.}
        & \textcolor{red!70!black}{Yes}\\

        \textbf{SQUID~\cite{yildiz2021persistent}}
        & SW & HE
        & \textcolor{red!70!black}{Model Drift.}
        & \textcolor{red!70!black}{Very Low (PF)}
        & \textcolor{green!55!black}{High}
        & \textcolor{red!70!black}{Low}
        & \textcolor{orange!80!black}{Const (Low Range)}
        & \textcolor{green!55!black}{Low}
        & \textcolor{green!55!black}{No} \\

        \rowcolor{black!3}
        \textbf{CHRT~\cite{dewinkel_ReliableTimekeepingIntermittent_2020}}
        & HW & Cap.
        & \textcolor{red!70!black}{Cap. Drift}
        & \textcolor{red!70!black}{Low}
        & \textcolor{green!55!black}{High}
        & \textcolor{green!55!black}{High}
        & \textcolor{red!70!black}{Linear}
        & \textcolor{red!70!black}{Scales w/ Cap.}
        & \textcolor{red!70!black}{Yes} \\

        \textbf{HARC~\cite{deep_HARCHeterogeneousArray_2020}}
        & HW & Cap.
        & \textcolor{red!70!black}{Cap. Drift}
        & \textcolor{red!70!black}{Low}
        & \textcolor{green!55!black}{High}
        & \textcolor{green!55!black}{High}
        & \textcolor{red!70!black}{Linear}
        & \textcolor{red!70!black}{Scales w/ Cap.}
        & \textcolor{red!70!black}{Yes} \\

        \rowcolor{black!3}
        \textbf{Fed.-Cap.~\cite{arreola2022federated}}
        & HW & Cap.
        & \textcolor{red!70!black}{Cap. Drift}
        & \textcolor{red!70!black}{Low}
        & \textcolor{green!55!black}{High}
        & \textcolor{green!55!black}{High}
        & \textcolor{red!70!black}{Linear}
        & \textcolor{red!70!black}{Scales w/ Cap.}
        & \textcolor{red!70!black}{Yes} \\

        \textbf{BIONIC~\cite{yildiz2026bionic}}
        & HW/SW & Cap./HE
        & \textcolor{red!70!black}{Cap. Drift}
        & \textcolor{red!70!black}{Very Low (PF)}
        & \textcolor{green!55!black}{High}
        & \textcolor{green!55!black}{High}
        & \textcolor{red!70!black}{Linear (w/ SR)}
        & \textcolor{green!55!black}{Low}
        & \textcolor{green!55!black}{No} \\

        \rowcolor{babyblueeyes!35}
        \textbf{\sysname (this work)}
        & HW/SW & MTJ
        &  \textcolor{green!55!black}{None}
        & \textcolor{green!55!black}{High}
        & \textcolor{green!55!black}{High}
        & \textcolor{green!55!black}{High}
        & \textcolor{green!55!black}{Constant}
        & \textcolor{green!55!black}{Low}
        & \textcolor{green!55!black}{No} \\

        \bottomrule
    \end{tabular}
    
    \begin{flushleft}
    \small
    \centering
    SRAM\,=\,SRAM remanence \quad Osc.\,=\,oscillator stabilization \quad Cap.\,=\,capacitor discharge \quad HE\,=\,harvested-energy modeling \\
    PF\,=\,power fluctuations \quad BCA\,=\,backup capacitor aging \quad ET\,=\,evironmental temperature \quad SR\,=\,sampling rate
    \end{flushleft}
    \label{tab:related-comparison}
\end{table*}

\noindpar{Range Analysis (\autoref{tab:cs_duration}, \autoref{fig:cs_duration}).}
\sysname tracks D4 at \qty{2.2}{\percent} and D5 at \qty{2.3}{\percent} geomean error. A capacitor tops out
near \qty{91}{\s}; reaching these ranges would require
\qtyrange{4.7}{65.2}{\milli\joule} per cycle and
\qtyrange{1.0}{14.6}{\milli\farad} of bulk capacitance for D4 and D5, respecitvely.

\subsection{Discussion and Limitations}


While \sysname's physics-based approach avoids the structural failure modes of capacitor clocks, several practical challenges remain.

\textbf{High reset voltage:} Reset voltage scales with array width due to the high resistance of MTJs. This can be mitigated by reducing device resistance (e.g., via area or tunnel barrier engineering) without affecting decay dynamics, changing squareness, or by redesigning the reset circuitry to operate on smaller sub-arrays. Alternative MTJ technologies (e.g., SOT or hybrid SOT-STT~\cite{kang_CriticalSwitchingCurrent_2021}) may further reduce switching requirements.

\textbf{Calibration overhead:} Calibration requires ADC sampling over the full decay interval (up to $\sim$\qty{1500}{\s} sampled every \qty{5}{\ms}). While this incurs time and energy overhead, it is a one-time process at setup and comparable to capacitor-based approaches. Future work may explore standardized calibration to reduce this cost.


\textbf{Device lifetime:}  MTJ endurance ($>10^{12}$ switching events) far exceeds timekeeping needs, and aging does not measurably alter switching behavior~\cite{deac2008bias,apalkov2016magnetoresistive}. The reset/measure cycle is the only wear path; at any plausible cycle rate the device outlives deployment.


\textbf{Fabrication challenges:} Scaling MTJ fabrication introduces variability and integration challenges. However, ongoing advances in MRAM technology continue to improve device uniformity and manufacturing, supporting the feasibility of large-scale deployment.

\textbf{Evaluation scope:} Exhaustive experimental design-space exploration is impractical: each array configuration 
(width, height, size, $E_b$, $R_p$, TMR)
requires a custom CMOS tape-out. Our design-space study therefore relies on simulation validated against fabricated devices (\autoref{sec:app-validation}). Area figures are estimates: capacitor sizes assume 0805 packages; \sysname's ${<}\qty{0.1}{\mm\squared}$ is a pessimistic upper bound.





\section{Related Works}


\subsection{Time-Sensitive Intermittent Computing}
Several software runtimes support time-sensitive execution in intermittent systems~\cite{hester2017timely,kortbeek2020time,yildiz2022immortal,erata2023etap,yildiz2023efficient,yildiz2024adaptable}, maintaining data freshness by restarting tasks when collected data becomes stale. Intermittent real-time scheduling approaches~\cite{maeng2020adaptive,islam2019zygarde,islam2020scheduling} similarly incorporate temporal constraints into task execution across power cycles. Both depend on accurate elapsed-time estimates across power failures, making persistent timekeeping a fundamental requirement.

\subsection{Timekeeping Approaches}
Prior work has explored time estimation without dedicated hardware by leveraging existing system components. TARDIS~\cite{hester2016persistent} estimates elapsed time from SRAM remanence, while oscillator-based approaches infer duration from stabilization behavior~\cite{alsubhi2020can}. SQUID~\cite{yildiz2021persistent} predicts off-time by modeling capacitor charging during active periods. More recently, BIONIC~\cite{yildiz2026bionic} estimates ambient power from capacitor charging times and predicts future charging delays, enabling adaptation to changing energy conditions without hardware modifications. However, its accuracy ultimately depends on the stability of harvested-power patterns and the validity of the underlying prediction model. Overall, these approaches either support only limited time ranges or rely on environment-dependent assumptions that can reduce robustness in changing conditions.

Capacitor-discharge techniques remain the dominant approach for long-range timekeeping. CUSTARD~\cite{hester2016persistent} uses a single capacitor, exposing a fundamental trade-off between startup cost and measurable duration: small capacitors charge quickly but support only short-range measurements, whereas larger capacitors extend range at the expense of energy overhead and startup latency. CHRT~\cite{dewinkel_ReliableTimekeepingIntermittent_2020,yildiz_TrackingTimeBetter_2024}, HARC~\cite{deep_HARCHeterogeneousArray_2020}, and federated capacitor designs~\cite{arreola2022federated} improve accuracy and range through multiple storage elements, but still inherit the same dependency on capacitor sizing. Consequently, measurement range remains coupled to energy consumption, area, and startup cost. Furthermore, because time estimation relies directly on capacitor behavior, accuracy is affected by capacitor aging and environmental variation over the deployment lifetime (\autoref{sec:results-aging}).

\autoref{tab:related-comparison} summarizes representative timekeeping approaches across key design axes. SRAM-remanence and oscillator-based techniques offer aging stability and low startup overhead, but are fundamentally limited in range. RTC-based solutions provide long-range, high-resolution timekeeping; however, like other capacitor-backed designs, they require deployment-time capacitor sizing based on expected harvesting conditions, potentially necessitating reconfiguration when those conditions change. Harvested-energy modeling approaches avoid explicit capacitor characterization but remain highly dependent on environmental conditions and often require retraining or recalibration. In contrast, FLINT achieves long-range, high-resolution, and environmentally robust timekeeping while maintaining constant energy cost, low startup overhead, and operation without hardware reconfiguration.

\subsection{Other MTJ Applications}
Outside of timekeeping, MTJs are most widely deployed as the storage element in STT-MRAM, where resistance states retain data non-volatilely with high endurance and radiation tolerance~\cite{ikegawa2021commercialization,wang2013low,marinella2021radiation}. They also serve as entropy sources: a fixed current across an STT-MTJ~\cite{singh2024cmos} or fixed voltage across a V-MTJ~\cite{shao2023probabilistic} drives 50\% switching probability, producing the random bitstreams required by p-bits~\cite{camsari2017stochastic,liu2025application} and p-dits~\cite{duffee2025p,duffee2026fap,duffee2026qap} used in probabilistic computing to solve combinatorial optimization problems~\cite{duffee2025integrated,borders2019integer}. Additionally, the sensitivity of the free layer to external magnetic fields enables use as magnetic field sensors~\cite{liu_ApplicationProbabilisticBits_2025}.

\section{Conclusion}

We present \sysname, the first timekeeper to read elapsed time from the physics of broken memory rather than from stored energy. Because an MTJ array's decay timescale is set by device geometry, \sysname's energy is independent of the duration it measures and its accuracy does not drift as the device ages. These are the two failure modes that limit capacitor clocks in long-lived deployments. \sysname tracks over 15 minutes of off-time within 10\% error while consuming \qty{1.03}{\micro\joule}, occupying under \qty{0.1}{\mm^2}, and adding \qty{50}{\us} of reset latency and \qty{70.83}{\us} to estimate time, and it extends to longer ranges at no added cost. Because idle arrays consume no power and a single chip can host many more arrays than any deployment needs, the same hardware covers short-range high-resolution tasks and long-range low-resolution ones without modification---a true ``one size fits all'' timekeeper, in contrast to capacitor clocks whose storage element must be sized at deployment time. Lightweight calibration and temperature correction hold this accuracy across realistic operating conditions, and over a one-year deployment \sysname makes 16--52\texttimes{} fewer scheduling errors than an aging capacitor clock.

\FloatBarrier

\begin{acks}

The authors would like to thank Andrew Liss (University of Michigan, Ann Arbor) for sharing his expertise on VLSI and integrated circuit design.

This research was partially supported by the U.S. National Science Foundation under award numbers CNS-2400463 and CCF-2322572. We would also like to acknowledge support by the Alfred P. Sloan Foundation, VMware, Google, and Catherine M. and James E. Allchin. Any opinions, findings, conclusions, or recommendations expressed in this material are those of the author(s) and do not necessarily reflect the views of the National Science Foundation or other supporters.

This research is supported in part through research cyberinfrastructure resources and services provided by the Partnership for an Advanced Computing Environment (PACE) at the Georgia Institute of Technology, Atlanta, Georgia, USA (RRID:SCR\_027619).

AI tools, including ChatGPT, Google Gemini, and Anthropic Claude (partially via GitHub Copilot), assisted with sections of this Work, including text, tables, graphs, and code. 
\end{acks}

\label{pg:last-page}

\printbibliography

\appendix

\setcounter{figure}{0}
\renewcommand{\thefigure}{A\arabic{figure}}

\clearpage

\section{Simulator Validation}\label{sec:app-validation}

Because the design-space exploration in this work relies heavily on simulation, we verify that the simulator reproduces the behavior of experimentally measured devices. \autoref{fig:hw_validation} compares the normalized array imbalance from the 21 fabricated devices against a Monte-Carlo simulation of a $64\times256$ \sysname array. Both datasets exhibit the same exponential decay, confirming that the simulator accurately captures the thermally activated Arrhenius switching dynamics $\tau \propto e^{E_b/k_bT}$ that govern \sysname operation. The curves begin to diverge after approximately \qty{20}{\s} because a finite ensemble of $N$ devices approaches a non-zero statistical imbalance floor of $\sqrt{2/\pi N}$ ($\approx0.17$ for $N{=}21$ versus $\approx0.006$ for the \num{16384}-device array). The plateau observed in experimental data therefore arises from finite-sample statistics rather than a systematic physical offset. This agreement between measurement and simulation provides confidence in the simulated design-space studies presented throughout this work.

\section{Further MTJ Quantification}\label{sec:app-more-mtj-exps}

\begin{figure}
    \centering
    \includegraphics[width=0.8\linewidth]{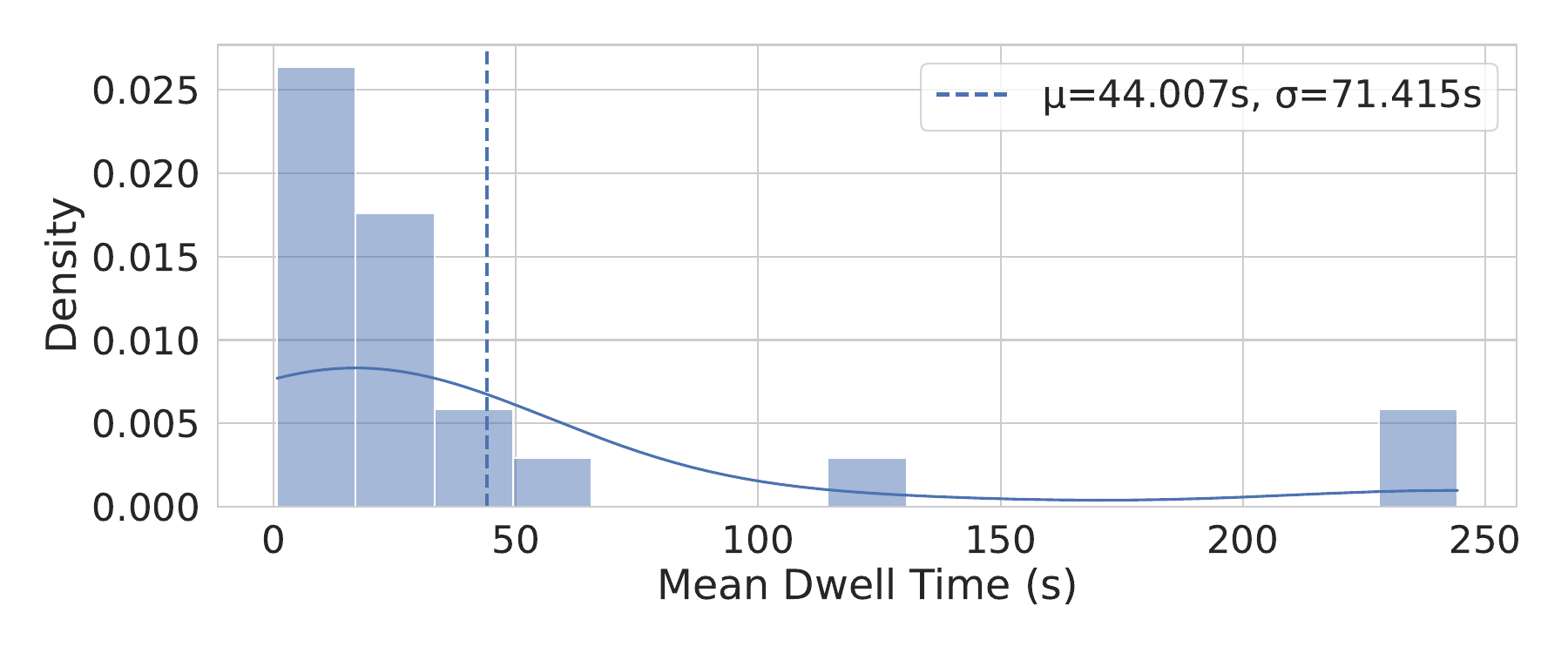}
    \Description{Histogram showing distribution of measured mean dwell times from 21 fabricated MTJ devices. Horizontal axis shows dwell time in seconds centered around 10 seconds, vertical axis shows device count. Distribution reflects device-to-device variation in thermal activation dynamics.}
    \caption{Mean dwell time distribution for 21 MTJ devices with a targeted \qty{10}{\s} dwell time.}
    \label{fig:dist_mean_dwell}
\end{figure}

\begin{figure}
    \centering
    \includegraphics[width=0.55\linewidth]{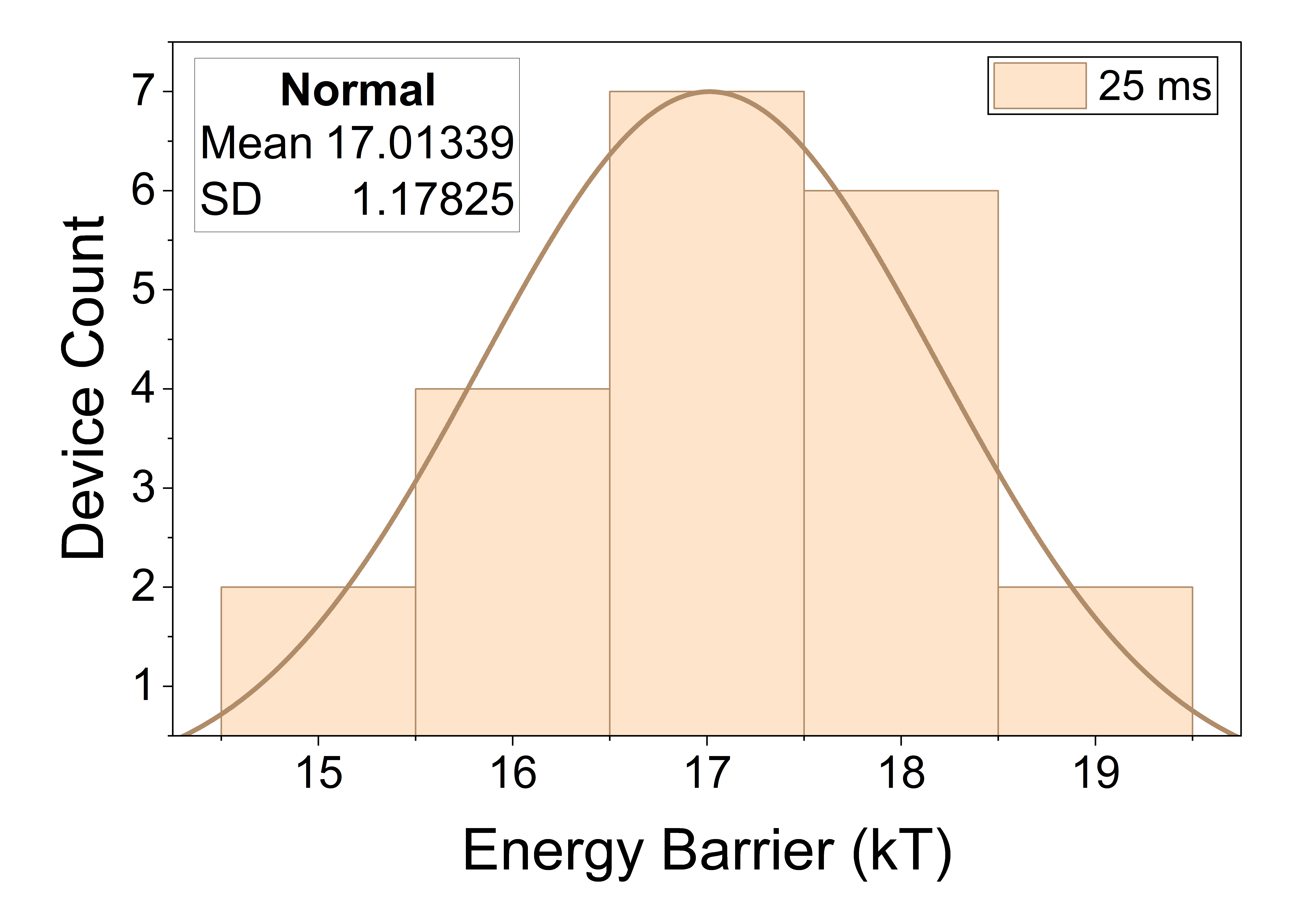}
    \Description{Histogram showing distribution of extracted energy barriers for 21 MTJ devices designed for 25 millisecond dwell time. Horizontal axis shows energy barrier in units of thermal energy, vertical axis shows device count. Normal distribution demonstrates predictability of energy barrier across devices.}
    \caption{Energy barrier distribution for 21 MTJ devices with a targeted \qty{25}{\ms} dwell time.}
    \label{fig:dist_eb_25ms}
\end{figure}

\autoref{fig:dist_mean_dwell} shows the distribution of mean dwell times for a targeted \qty{10}{\s} dwell time, reflecting the raw dwell times before conversion to energy barriers via the N\'eel-Arrhenius relation. \autoref{fig:dist_eb_25ms} shows the energy barrier distribution for devices targeting a \qty{25}{\ms} dwell time. Together they demonstrate the tunability of the mean energy barrier: the normal distribution shifts to smaller energy barriers as the targeted dwell time decreases, and we expect a similarly predictable shift at other dwell-time targets.

\section{Further Single Array Simulation Results}\label{sec:app-more-sims}

\begin{figure}[htbp]
    \centering
    \includegraphics[width=0.9\linewidth]{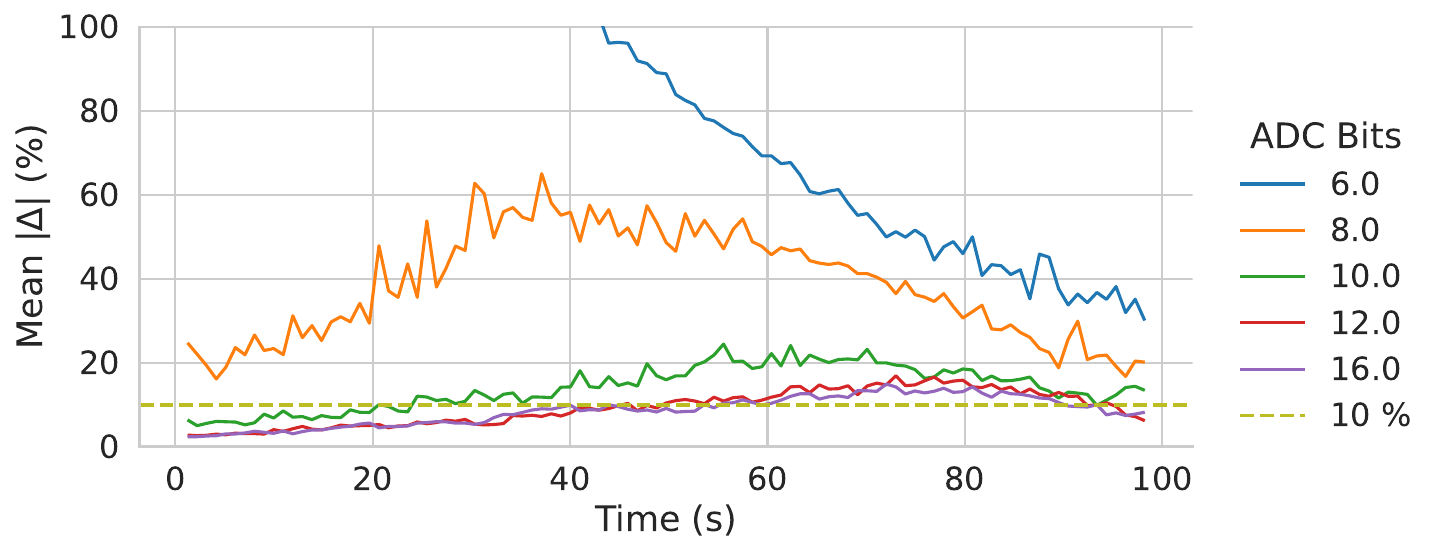}
    \Description{Line plot showing estimation error percentage versus elapsed time. Multiple colored lines represent different ADC resolutions from 8 to 16 bits. Error decreases substantially up to 12 bits, then improvements plateau.}
    \caption{Timekeeper error with varying numbers of ADC bits.}
    \label{fig:eb23_adc_bits}
\end{figure}

\noindpar{ADC bit selection.} In \autoref{fig:eb23_adc_bits}, we plot the timekeeper error with a varying number of ADC bits. We find that at least 10 bits are required to achieve timekeeping accuracy of at least 10\%. For all further experiments, we choose a 12-bit ADC because it is offered by many low-power microcontrollers~\cite{texasinstruments_MSP430FR596xMSP430FR594xMixedSignal_2018, espressifsystems_ESP32WROOM32Datasheet_2025}.

\begin{figure}[htbp]
    \centering
    \includegraphics[width=\linewidth]{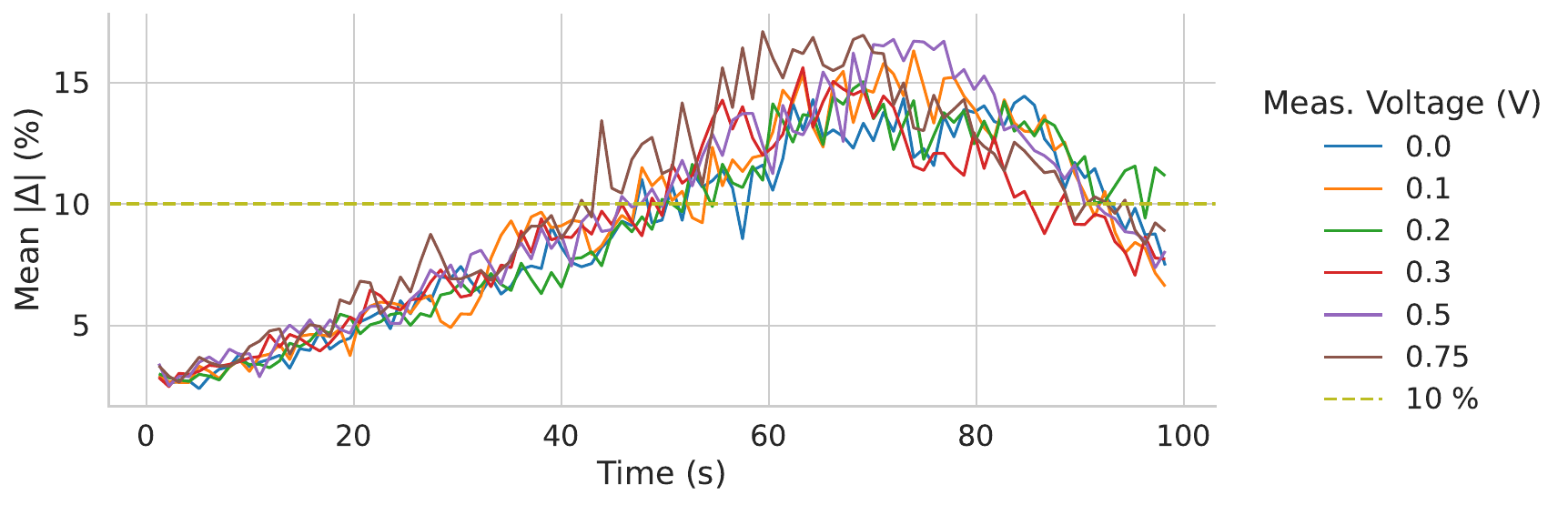}
    \Description{Line plot showing estimation error percentage versus elapsed time. Multiple colored lines represent measurement voltages from 0.1 to 0.5 volts across MTJ devices. Calibration enables acceptable error below 10 percent for all voltages up to 0.5 volts.}
    \caption{Timekeeper error with varying measurement voltages (with calibration).}
    \label{fig:eb23_v_meas}
\end{figure}

\noindpar{Impact of Measurement Voltage.} Finally, the voltage across the MTJs, among other intrinsic factors, determines the TMR ratio of the device. TMR drops when the voltage across the device increases, affecting the sensitivity of time estimation negatively; it is desirable to operate under low voltage conditions. We plot the timekeeper error as a function of increasing device voltages and appropriate prior calibration in \autoref{fig:eb23_v_meas}. Up to a range of 0.5V per device, our calibration is able to achieve a 40-second range under 10\% error. As noted prior, \sysname is designed to operate at 0.1\,V across each device, keeping our operating point well under the upper limit and enabling larger TMR ratios and better time resolution.

\end{document}